\documentclass[aps,twocolumn,pre]{revtex4-2}

\usepackage{amsmath}
\usepackage{amssymb}
\usepackage{comment}
\usepackage{graphicx}
\usepackage{xcolor}
\usepackage{amsmath}
\usepackage{hyperref}% add hypertext capabilities

\newcommand{\xst}{x_\text{\tiny{st}}}
\newcommand{\mst}{m_\text{\tiny{st}}}

\newcommand{\ac}{a_\text{c}}
\newcommand{\ts}{\tau^*}
\newcommand{\qi}{q_\infty}

\begin{document}
 
\title{Complete aging in the noisy voter model enhances consensus formation}

\author{
Jaume Llabr\'es$^1$,
Sara Oliver-Bonafoux$^1$, 
Celia Anteneodo$^{2,3}$, 
Ra\'ul Toral$^1$
}
\affiliation{$^1$Institute for Cross-Disciplinary Physics and Complex Systems IFISC (CSIC-UIB), Campus UIB, 07122 Palma de Mallorca, Spain,\\
$^2$Department of Physics, Pontifical Catholic University of Rio de Janeiro, PUC-Rio, Rua Marquês de São Vicente, 225, 22451-900 Rio de Janeiro, Brazil\\
$^3$National Institute of Science and Technology for Complex Systems, INCT-SC, Rio de Janeiro, Brazil
}

%\email{jaumellabres@ifisc.uib-csic.es}
%\email{saraoliver@ifisc.uib-csic.es} 
%\email{celia.fis@puc-rio.br}%
%\email{raul@ifisc.uib-csic.es}

%\date{\today}% 

\begin{abstract}
%We analyze the effects of inertia or resilience to change (aging) in the noisy voter model. In the noisy version, besides the social rule of simple contagion, opinion can change randomly, regardless of the interactions. To take inertia into account, we consider that the probability of change decays algebraically with age $\tau$, with a characteristic time $\tau^*$. We investigate the situation where aging acts on both socially influenced and random opinion changes (complete aging), and compare it with previous results where aging acts only on pairwise interactions (partial aging). We obtain analytical predictions for the dynamics in the mean-field limit, in excellent agreement with agent-based simulations on a complete graph. We observe that order (with the predominance of one of the opinions) is favored by an optimal value of $\tau^*$, even more sharply when aging is complete. However, when the aging probability decays asymptotically to zero for large $\tau$, a steady state is not always attained for complete aging. 

We investigate the effects of aging in the noisy voter model considering that the probability to change states decays algebraically with age $\tau$, defined as the time elapsed since adopting the current state. We study the complete aging scenario, which incorporates aging to both mechanisms of interaction: herding and idiosyncratic behavior, and compare it with the partial aging case, where aging affects only the herding mechanism. Analytical mean-field equations are derived, finding excellent agreement with agent-based simulations on a complete graph. We observe that complete aging enhances consensus formation, shifting the critical point to higher values compared to the partial aging case. However, when the aging probability decays asymptotically to zero for large $\tau$, a steady state is not always attained for complete aging.

%\JL{In the limit where the aging kernel decays asymptotically to zero, the system exhibits a unique dynamical regime, undergoing a first-order-like transition between ordered and disordered states.}
\end{abstract}

\maketitle

\section{Introduction}
\label{sec:intro}
The dynamics of opinion formation in social systems is a rich and complex subject, where various mechanisms influence how individuals update their beliefs and how these updates aggregate to form collective behaviors. The tools of statistical physics offer valuable insight into this phenomenon, providing a natural framework for analyzing complex systems with many interacting components~\cite{Castellano:2009, Conte:2012,Galam:2012,SanMiguel2020b}. The dynamics of opinion formation can then be studied using concepts like phase transitions, criticality, and symmetry breaking. 

In many social models, individuals are assumed to update their opinions by imitating others in their social network. However, this imitation process is often imperfect, and the interplay of noise, arising from factors such as independent choices or stochastic perturbations, can profoundly affect the system's collective behavior. Hence, one fundamental aspect of the opinion formation process is the competition between the drive toward uniformity (understood as consensus in this context), which arises from social contagion and influence, and the inherent diversity introduced by individual noise or random factors. 

While traditional models of opinion dynamics, such as the voter model~\cite{Krapivsky1992,liggett2013,Mobilia:2003rr,2007Mobilia}, focus on deterministic rules of imitation, the incorporation of noise introduces stochasticity into the system, leading to a variety of possible outcomes. These noisy dynamics can result in the persistence of diversity in a population or, conversely, can promote the emergence of consensus depending on the strength of social influence and the level of randomness. Several noisy opinion dynamics models have been developed to capture this phenomenon, including the noisy voter model~\cite{Kirman1993, Granovsky:1995,carro2016noisy, peralta2018stochastic, khalil2018zealots}, the noisy threshold voter model~\cite{SWeron2018, Peralta2020}, and noisy kinetic-exchange models~\cite{Crokidakis2014, Vieira2016}. In these models, the social influence rule is typically modified to include a probabilistic element, such that with a probability $1-a$, individuals update their opinion by following the majority, while with probability $a$, individuals change their opinion independently of others, introducing a form of random noise. The parameter $a$, often referred to as the ``social temperature," controls the strength of this randomness and therefore the system's capacity to reach consensus. When $a$ is small, the system is more likely to exhibit coherent behavior and reach a state of collective order, while larger values of $a$ hinder consensus formation, making it more difficult for the system to settle into a common majority opinion. 

In addition to the noisy dynamics of opinion formation, recent studies have highlighted the importance of non-Markovian effects that extend the traditional assumptions of the voter model. One such effect is ``latency time'', where agents remain inactive for a period after changing their opinion, effectively introducing a delay in their ability to engage in further social interactions. This delay can be thought of as a memory effect, where past interactions influence future behavior, even in the absence of immediate social influence. Palermo et al.~\cite{Palermo:24} have explored this phenomenon, showing that latency can significantly alter the dynamics of opinion formation, slowing down the process of consensus or fostering persistent disagreement in the system. Another key non-Markovian effect is ``persistence'', where agents, after adopting a particular opinion, become ``transient zealots"~\cite{arenzon2022, arenzon2024}. These agents exhibit a strong attachment to their opinion, resisting social influence and thereby breaking the symmetry of the opinion dynamics. This persistence can act as a stabilizing mechanism for a minority opinion, preventing it from being absorbed into the majority, and can also delay or prevent the system from reaching consensus.

Another non-Markovian effect that influences the dynamics of opinion formation is that of ``aging'', which refers to the tendency of individuals to become less susceptible to changing their opinions the longer they hold a particular belief. Initially introduced as ``inertia"~\cite{Stark2008}, aging has been shown to slow down the dynamics of opinion change, potentially affecting both the time to reach consensus and the final state of the system. Interestingly, while aging typically acts as a stabilizing factor in the system, recent studies have demonstrated that the slowing down of microscopic dynamics induced by aging can paradoxically speed up the time needed for the system to reach a macroscopically ordered consensus~\cite{Stark2008}. This counterintuitive result suggests that the dynamics of opinion formation are not solely governed by the local interactions between agents, but also by the underlying time-dependent properties of the system, which can shift the system's behavior in nontrivial ways.

In its simplest form, aging extends the standard voter model by introducing time-dependent transition rates that depend on the duration an agent has spent in a given opinion state. In particular, it is considered that the transition rates for flipping opinions monotonously decay as an agent spends more time in a particular opinion state. This aging effect reduces the probability of switching opinions over time, reflecting the increased resistance to change that naturally arises in social systems as individuals become more rooted in their beliefs.

While previous studies~\cite{Artime2019,VieiraPRE2024} including our own work~\cite{ourPhysics2024}, have explored the case where aging only affects social contagion, a situation that we denote here by ``partial aging'', we now turn our attention to the more general scenario of ``complete aging'', where aging is applied not only to the social contagion mechanism (the imitation of neighbors' opinions) but also to the random, noisy component of the opinion dynamics. This extension allows us to explore how the influence of aging on both social contagion and independent opinion change affects the system's behavior. Specifically, we seek to compare the effects of partial and complete aging on the noisy voter model, examining whether the different aging schemes lead to qualitatively different behaviors in terms of consensus formation, the time to reach equilibrium, and the stability of minority opinions. By comparing these different aging scenarios within the framework of the noisy voter model, we can identify the critical factors that determine whether a social system will evolve toward consensus or remain in a state of persistent disagreement, shedding light on the role of individual resistance to opinion change in complex social systems.

The structure of the paper is organized as follows: In Sec.~\ref{sec:model}, we describe the updating rules governing the stochastic process. In Sec.~\ref{sec:MF_description}, a mean-field description of the system under an adiabatic approximation is provided and the particular form of aging considered is specified. The main results of our research are presented in Sec.~\ref{sec:results}, which is divided into two blocks. In Sec.~\ref{sec:qinf>0}, a comparative study between partial and complete aging is performed with analytical results supported by numerical simulations on complete graphs. A particular case is explored in Sec.~\ref{sec:qinf0}, where the adiabatic approximation breaks down, revealing a distinct dynamical regime. Finally, concluding remarks and perspectives for future research are discussed in Sec.~\ref{sec:conclusions}. The appendices contain the more technical details of the calculations and some additional figures.

\section{The model: noisy voter with aging}\label{sec:model}
 
Consider a set of $N$ agents, each of them holding a binary state variable $s_i\in\{-1,+1\},\,i=1,\dots,N$, representing the position of agent $i$ against or in favor of the topic under discussion. Agents can change their state due to either social influence---by a non-reciprocal pairwise interaction of simple contagion---or randomly, otherwise known as idiosyncratic changes. In addition, each agent $i$ holds an internal age variable $\tau_i=0,1,2,\dots$, representing the number of consecutive failed attempts to change. 

Let us describe explicitly the evolution rules of the model. Initially, each state variable $s_i$ takes a random value and all internal times $\tau_i$ are set to zero. Then,
\begin{enumerate}
\item An agent $i$ is randomly selected.
\item With probability $(1-a)$, the social rule is chosen and activated with probability $q(\tau_i)$. In that case, agent $i$ copies the state of a randomly chosen neighbor $j$, $s_i \to s_j$.
\item Otherwise, with probability $a$, the idiosyncratic rule is chosen and activated with probability $\tilde{q}(\tau_i)$. If this happens, the state $s_i$ randomly takes one of the two possible values.

\item Regardless of the update mechanism used by agent~$i$, its internal age is updated as follows: If the state $s_i$ changes, the age is reset to 0, i.e., $\tau_i \to 0$. Otherwise, the age is incremented by one unit, i.e., $\tau_i \to \tau_i + 1$. 
\end{enumerate}
Time is measured in Monte Carlo steps (MCS), so that one unit of time corresponds to $N$ repetitions of this process. Thus, on average, each agent is selected once per MCS.

The functions $q(\tau)$ and $\tilde{q}(\tau)$ represent the resistance of individuals to accept the changes arising from herding or idiosyncratic behavior, respectively, as a function of their age $\tau$. From this point on, we adopt the notation $q_\tau \equiv q(\tau)$ and $\tilde{q}_\tau \equiv \tilde{q}(\tau)$ for simplicity.

In the next section, we present a mean-field description of the noisy voter model for generic forms of the aging kernels $q_\tau$ and $\tilde{q}_\tau$.

\section{Mean-field description}\label{sec:MF_description}

We consider the mean-field scenario, or all-to-all coupling, where every individual is a neighbor of any other individual. Let us denote by $x^+_\tau$ the fraction of agents in state $+1$ and age $\tau$, and by $x = \sum_{\tau = 0}^\infty x^+_\tau$ the corresponding global fraction. As detailed in Appendix~\ref{app:rates}, it is possible to obtain rate equations providing the dynamical evolution of $x^+_\tau(t)$ and $x^+(t)$. Under the assumption that the microscopic variables $x^+_\tau$ quickly reach the steady state and can be adiabatically eliminated, one obtains the following closed evolution equation for the global variable~$x$ (see Appendix~\ref{app:rates} for details)
\begin{align}\label{eq:dxdtVgeneral}
\frac{dx}{dt} &=G(x)\equiv \frac{a}{2} \left[(1-x)\tilde\Phi(x) - x\tilde\Phi(1-x)\right] \nonumber\\
&+ (1-a) x (1-x) \left[\Phi(x)- \Phi(1-x)\right],
\end{align}
where the functions $\Phi(x)$, $\tilde{\Phi}(x)$ are given by

\begin{equation} \label{eq:Phis}
\Phi(x) = 
\frac{\sum_{\tau=0}^\infty q_\tau F_\tau(x)}{\sum_{\tau=0}^\infty F_\tau(x)} \, , \,
\tilde\Phi(x) = 
\frac{\sum_{\tau=0}^\infty \tilde{q}_\tau F_\tau(x)}{\sum_{\tau=0}^\infty F_\tau(x)} ,
\end{equation}
with
\begin{equation} \label{eq:Ftau}
F_0(x)=1, \hspace{1cm}
F_\tau(x)\equiv \prod_{k=0}^{\tau-1}\Lambda(q_k\, x,\tilde{q}_k,a),\quad \tau\ge 1,
\end{equation}
and
\begin{equation} \label{eq:Lambda_def}
\Lambda(z,q,a)= a\left(1-\frac{q}{2} \right) +(1-a)(1-z).
\end{equation}

The steady-state solutions, $\xst$, are found by setting $G(x) = 0$ in Eq.~\eqref{eq:dxdtVgeneral}. From the structure of Eq.~\eqref{eq:dxdtVgeneral}, it is clear that $\xst = 1/2$ is a fixed point of this dynamical system. This solution corresponds to the scenario in which the population is evenly split, with exactly half holding each of the two opinions. This corresponds to a polarized society or, from the viewpoint of statistical physics, a disordered state. Moreover, the invariance of Eq.~\eqref{eq:dxdtVgeneral} under the transformation $x \to 1 - x$ implies that any additional solutions must occur in symmetric pairs, $\xst$ and $1-\xst$. Hence, we consider, without loss of generality, $1/2\le\xst\le 1$. The stability of these fixed points is determined by the sign of the derivative of $G(x)$ evaluated at each solution, i.e., $dG(x)/dx|_{x=\xst}$. 

As long as $\xst>1/2$, the $+1$ opinion will be held by a majority of the population, a situation that is recognized as consensus or order. Our primary interest lies in identifying non-trivial steady-state solutions of Eq.~\eqref{eq:dxdtVgeneral} and investigating the transitions between these solutions and the disordered one that arise as the model parameters are varied. 

By setting $\tilde{q}_\tau=q_\tau=1$, which implies $\tilde{\Phi}(x)=\Phi(x)=1$, Eq.~\eqref{eq:dxdtVgeneral} becomes the rate equation of the noisy voter model without aging~\cite{Kirman1993}. In this case, $\xst=1/2$ is the only fixed point, which is stable and no transitions are possible.

When considering aging effects, several functional forms for the activation probability have been explored in the literature. In this work, we adopt a rational function of the age
\begin{equation}
 q_\tau=\frac{\qi\tau+q_0\tau^*}{\tau+\tau^*},
\end{equation}
where $q_0, \qi \in[0,1]$ denote the initial and asymptotic ($\tau\to\infty$) values of the function, respectively, and $\tau^*>0$ characterizes the rate of change of the aging kernel, such that the larger $\tau^*$, the slower the function $q_\tau$. If $\qi<q_0$, the function $q_\tau$ decreases with age, a behavior commonly referred to as aging. On the other hand, if $\qi>q_0$, the function $q_\tau$ increases with age, a phenomenon known as ``anti-aging"~\cite{Peralta2020a}. This latter case lies beyond the scope of this paper and we focus exclusively on the aging scenario. 

Most of the works in the literature that have considered this functional form of aging have fixed the values of the parameters to $\qi=0$, $q_0=1/2$, $\tau^*=2$~\cite{Artime2018, Artime2019, Abella2022, Abella2023}. Others have considered $\qi=0$, $q_0=1$, and general values of~$\tau^*$~\cite{VieiraPRE2024, ourPhysics2024}. 
A detailed study in terms of all parameters is performed in the case of the (noise-less) voter model in Ref.~\cite{Peralta2020a}.

In this work, we make two primary contributions. First, we extend the analysis of the noisy voter model with partial aging presented in Ref.~\cite{ourPhysics2024} by incorporating the dependence on the asymptotic value $\qi$ using the following scheme
\begin{eqnarray}\label{eq:qtau-partial}
q_\tau &=& \frac{\qi\tau+\tau^*}{\tau+\tau^*},\\
\tilde{q}_\tau &=& 1.
\end{eqnarray}
Second, we explore the complete aging scenario, in which we also include aging effects in the idiosyncratic behavior, assuming that the same aging profile governs both mechanisms, namely 
\begin{eqnarray}\label{eq:qtau}
q_\tau =\tilde{q}_\tau &=& \frac{\qi\tau+\tau^*}{\tau+\tau^*}.
\end{eqnarray}
In both cases, we have set $q_0 =1$, while $\qi\in[0,1]$. This assumption is made without loss of generality, as a different value for $q_0$ can be reabsorbed in a global time scale factor, which is irrelevant for the study of the steady-state solutions.

The scenario in which aging affects only the idiosyncratic mechanism is not considered here, as it simply reproduces the phenomenology of the noisy voter model without aging: the disordered state is the only solution and no phase transition is possible. 

\begin{figure}[b!] 
 \centering
\includegraphics[width=0.4\textwidth]{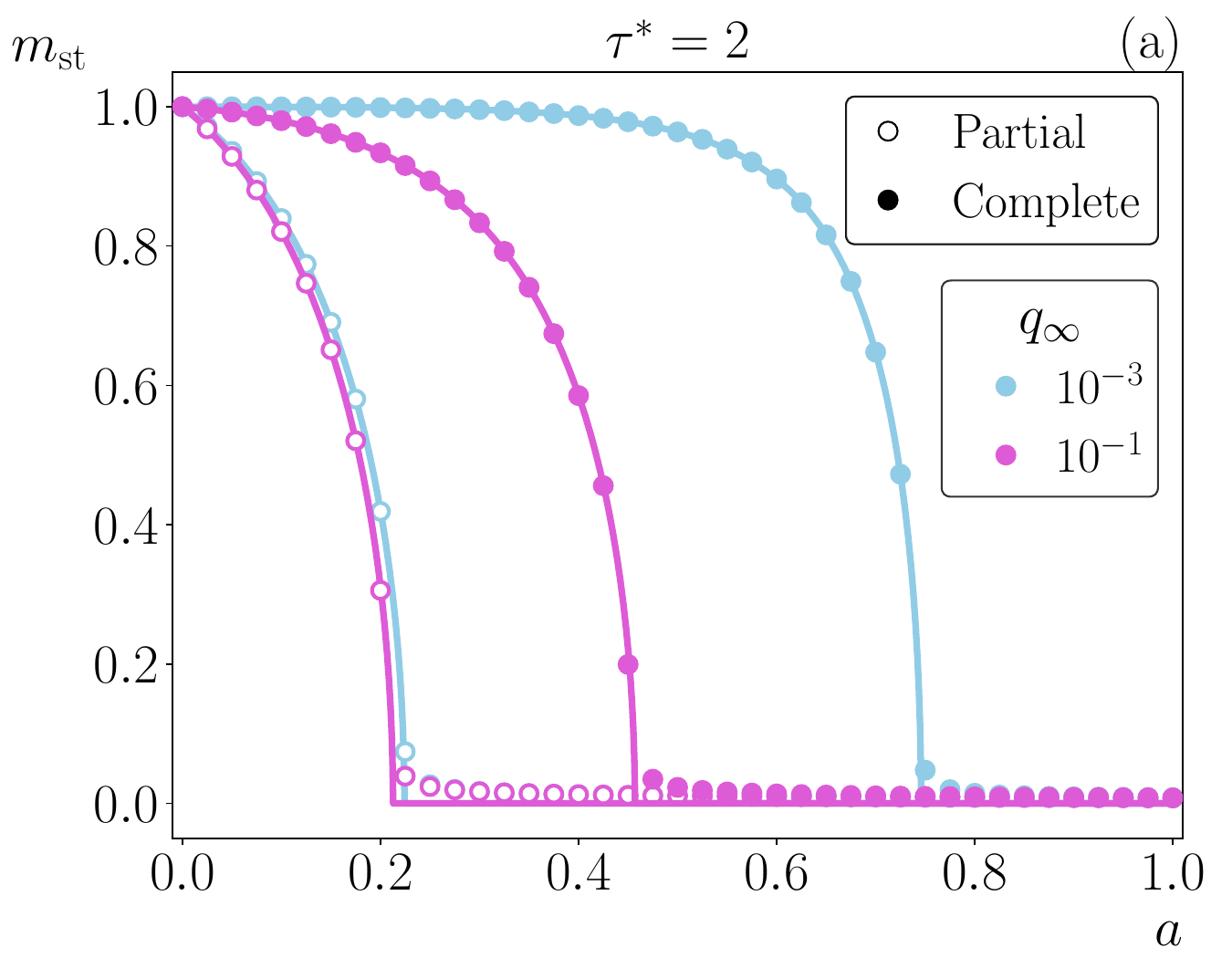}
\includegraphics[width=0.4\textwidth]{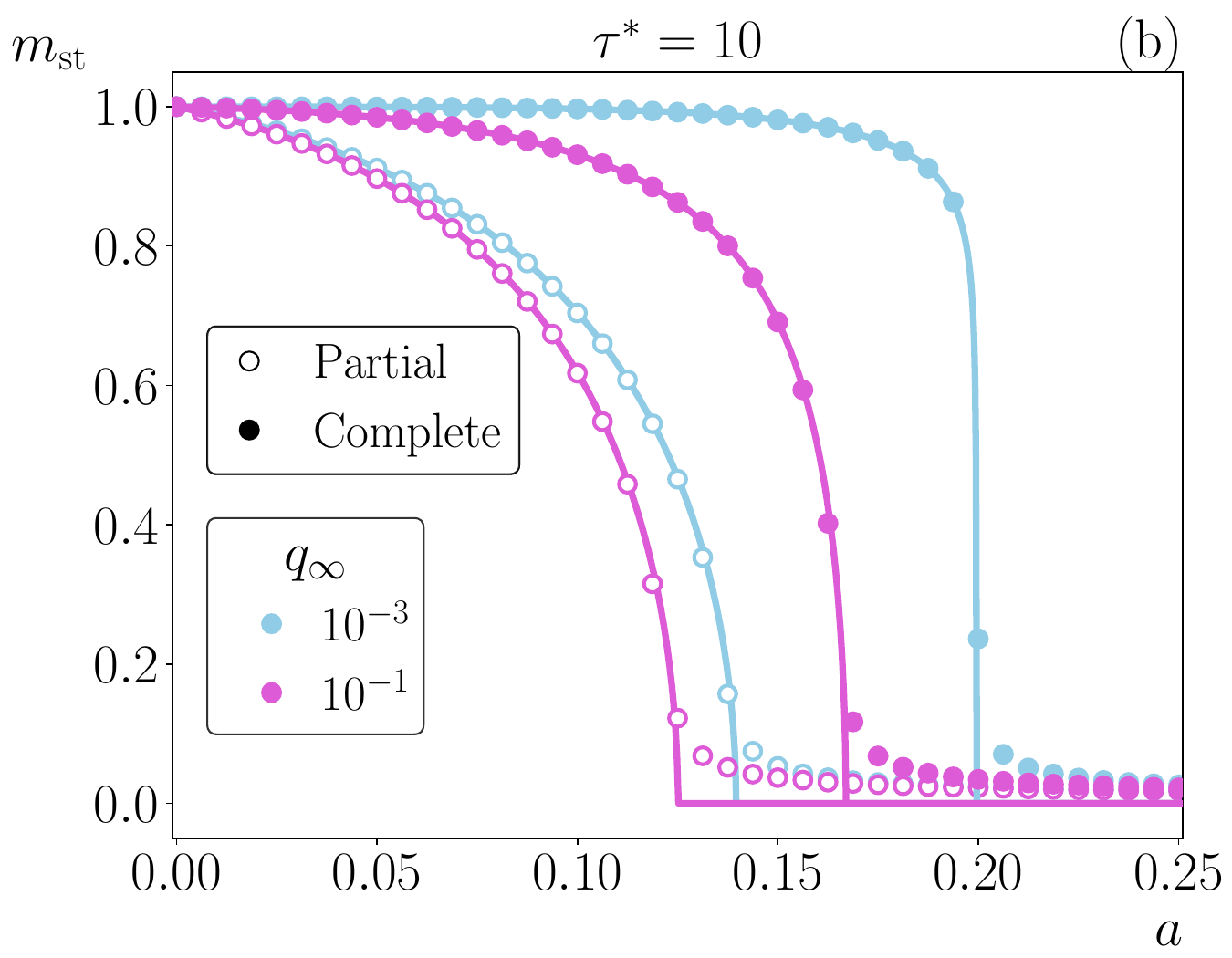}
 \caption{
 Magnetization, $\mst$, versus the noise intensity, $a$, for two values of $\qi$ and for (a) $\tau^*=2$ and (b) $\tau^*=10$. Solid lines correspond to the numerical solution of the stationary form of Eq.~\eqref{eq:dxdtVgeneral}, while symbols represent the results of numerical simulations of the agent-based model described in Sec.~\ref{sec:model}. In the simulations we have considered a population of $N=10^4$ agents, and averaged the magnetization over $10^6$~MCS after a transient of $10^6$ MCS.}
 \label{fig:m_vs_a_qinf>0}
\end{figure}

%%%%%%%%%%%%%%%%%%%%%%%%%%%
\section{Results} \label{sec:results}
In this section, we present analytical results for the noisy voter model with both partial and complete aging, complemented by numerical simulations on complete graphs. The section is organized into two parts. 

The first part focuses on comparing the partial and complete aging scenarios for $\qi > 0$. We conduct a standard stability analysis of the fixed points, examine the phase transition between coexistence and consensus states, and analyze the agents' age distribution. 

The second part explores the special case $\qi = 0$. While the partial aging scenario shows no significant deviations from the $\qi > 0$ case, the complete aging scenario presents a unique phenomenology which requires a more detailed investigation. This analysis is complicated by mathematical difficulties related to the convergence of the series appearing in Eq.~(\ref{eq:Phis}).

\subsection{Comparison of partial and complete aging for $\qi>0$}
\label{sec:qinf>0}
For both types of aging, partial and complete, a second order phase transition from an ordered to a disordered state occurs at a critical value of the noise intensity, denoted as $\ac$. For $a<\ac$, Eq.~\eqref{eq:dxdtVgeneral} has a pair of stable symmetric solutions $\xst$, $1-\xst$, while the disordered solution $\xst=1/2$ is unstable. Although it is not possible to obtain an explicit expression for these nontrivial solutions, they can be determined numerically with high accuracy. For $a>\ac$, on the other hand, only the solution $\xst=1/2$ exists and it is stable. 

In Fig.~\ref{fig:m_vs_a_qinf>0}, we plot the stationary magnetization, \makebox{$\mst=|2x_\text{\tiny{st}}-1|$}, as a function of the noise intensity, $a$, for different values of $\ts$ and $\qi$. In the figure we include, in addition to the analytical results, the outcomes of numerical simulations with $N = 10^4$ agents. The strong agreement between the simulations and the analytical results supports the validity of the adiabatic approximation introduced in the theoretical treatment. However, some discrepancies are observed for noise intensity values close to the critical point $\ac$. These discrepancies arise from the finite number of agents considered in the numerical simulations, which contrasts with the thermodynamic limit $N \to \infty$ assumed in the theoretical framework. The figure shows that the main effect of complete aging is to increase the consensus region by shifting the critical point~$\ac$ to higher values. Furthermore, the case of complete aging is more sensitive to the asymptotic level~$\qi$, as can be observed for both values of $\ts$ displayed in the figure. 

\begin{figure}[b!] 
\includegraphics[width=0.45\textwidth]{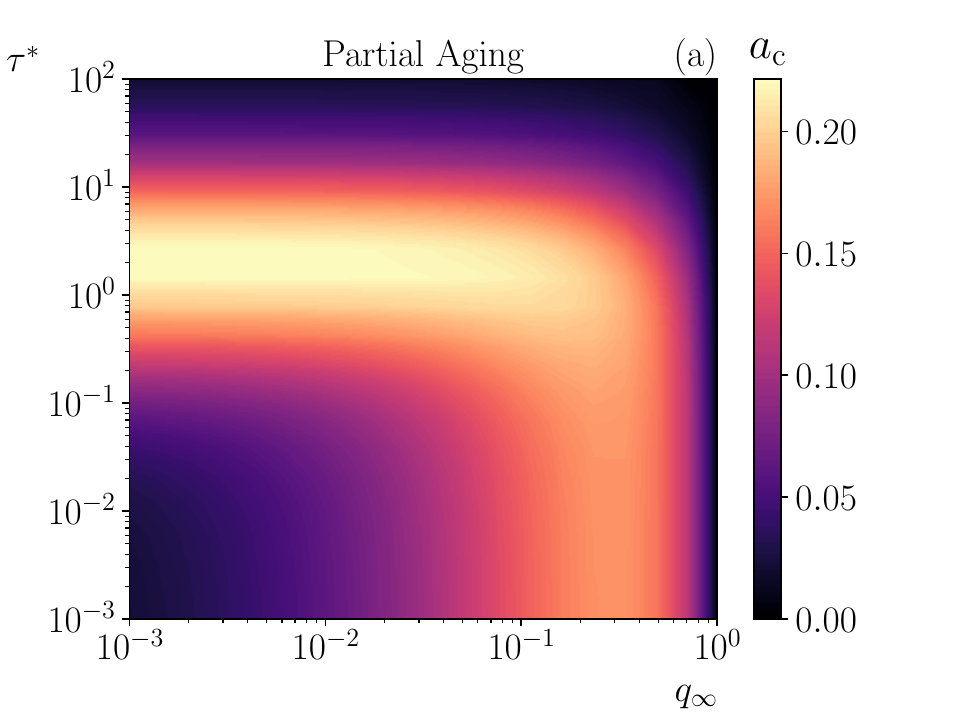}
\includegraphics[width=0.45\textwidth]{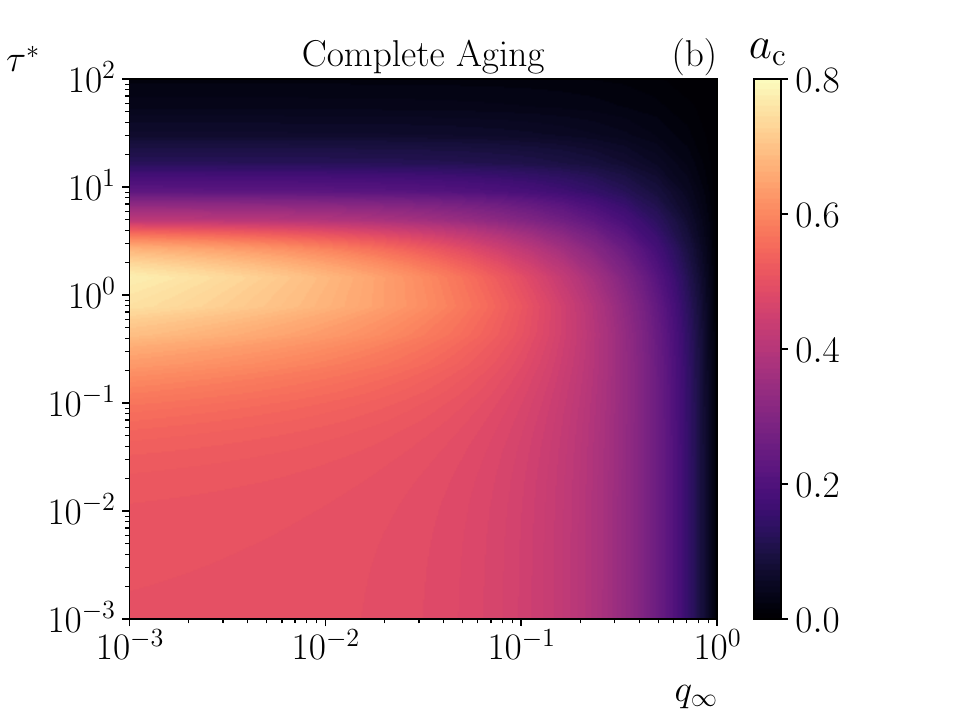}
\caption{Contour plot of the critical value $\ac$ in the parameter space $(\qi,\tau^*)$, in the case of (a) partial aging and (b)~complete aging.
The results are derived from the numerical solution of Eqs.~\eqref{eq:ac_partial} and~\eqref{eq:ac_complete}, respectively.}
\label{fig:acs_heat}
\end{figure}

\begin{figure}[b!] 
\includegraphics[width=0.38\textwidth]{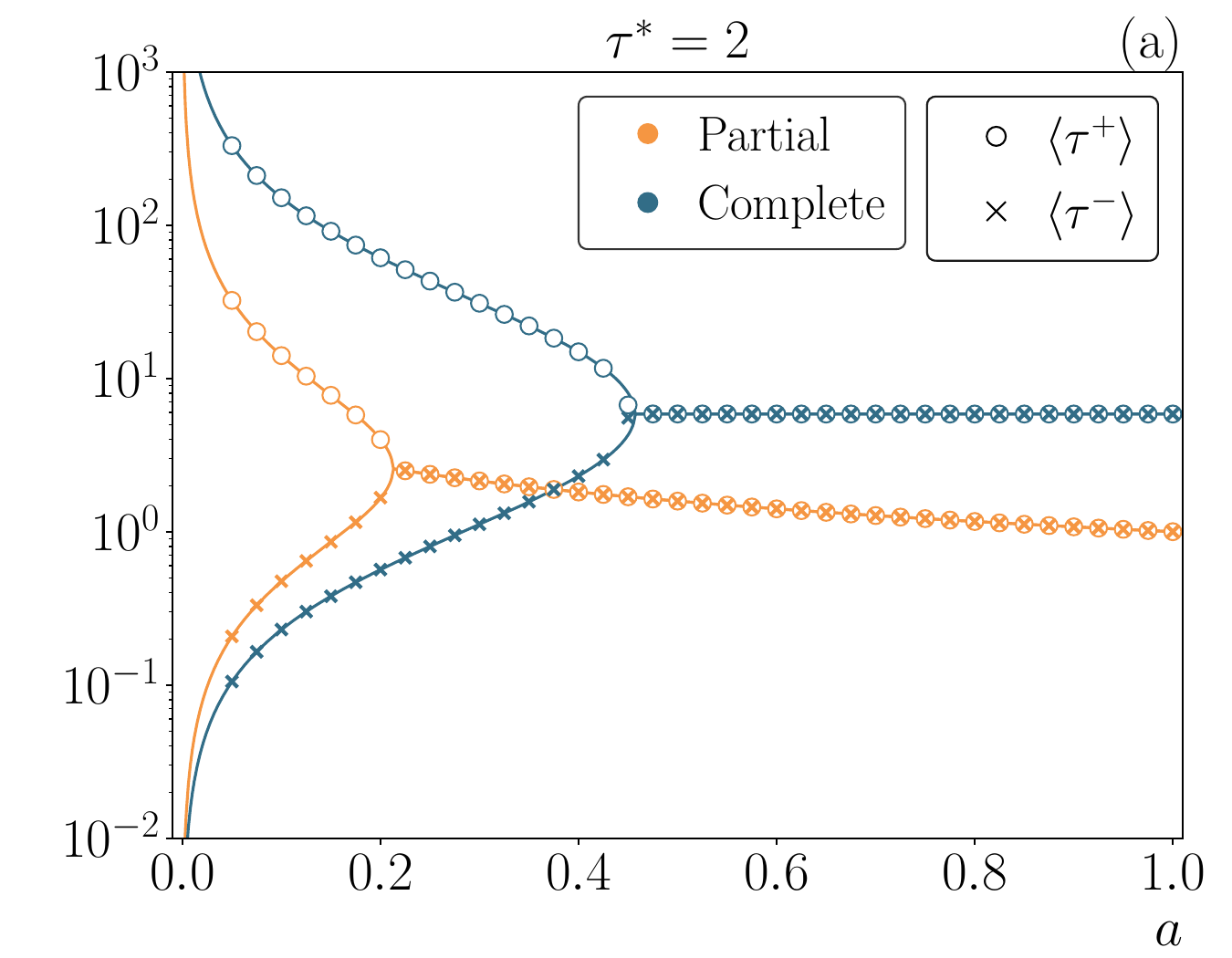}
\includegraphics[width=0.38\textwidth]{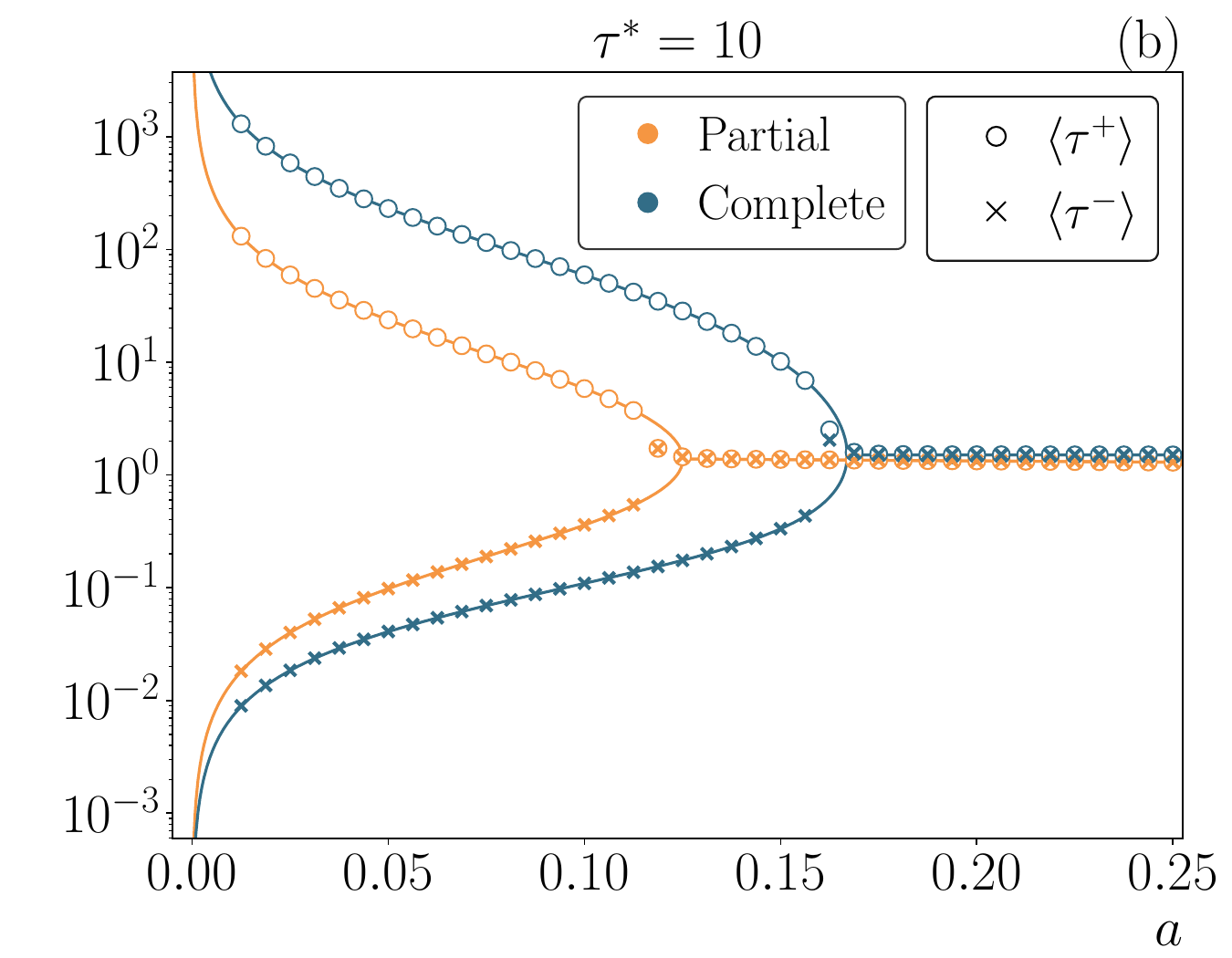}
 \caption{Mean internal age $\langle \tau^s\rangle$ of agents in state $s = \pm 1$ versus the noise intensity $a$, considering that the majority is in state $+1$. We have set $\qi = 10^{-1}$ and (a) $\ts = 2$ and (b) $\ts = 10$. Symbols correspond to the results obtained in numerical simulations with $N = 10^4$ agents, in which we have averaged over $10^6$ MCS after a transient of $10^6$ MCS. On the other hand, solid lines correspond to the theoretical solutions given by Eqs.~(\ref{eq:app:mean_taus},\ref{eq:app:general_T}).}
 \label{fig:taus_qinf>0}
\end{figure}

Unfortunately, it is not possible to obtain an analytical expression for the critical point $\ac$. However, it can be determined numerically by identifying the point at which the trivial solution $\xst=1/2$ changes its stability. This is achieved by solving the equation \makebox{$dG(x)/dx|_{x=1/2}=0$}, which leads to the relation~\cite{ourPhysics2024}
\begin{align} \label{eq:ac_partial}
 (1-\ac)\Phi'(x)|_{x=1/2} = 2 \ac,
\end{align}
in the partial aging case, and to the relation
\begin{align} \label{eq:ac_complete}
 \Phi'(x)|_{x=1/2} = 2 \ac\Phi(x)|_{x=1/2},
\end{align} 
in the complete aging case. Note that in both Eqs.~\eqref{eq:ac_partial} and~\eqref{eq:ac_complete}, the function $\Phi(x)$ also depends on the critical value $\ac$. By numerically solving these equations, we can plot the critical value $\ac$ in the parameter space $(\qi, \ts)$ for both partial and complete aging, as shown in Fig.~\ref{fig:acs_heat}. Note that in this figure brighter regions indicate higher values of $\ac$, and that the vertical $\ac$ scales vary significantly in the cases of partial and complete aging. For any combination of $\qi$ and $\ts$ values, complete aging yields a higher critical point $\ac$ than partial aging, illustrating that complete aging enhances consensus among the population. While the maximum values of $\ac$ are slightly higher than 0.2 for partial aging, with complete aging we reach critical point values near 0.8. The maximum values of $\ac$ occur for $\ts \sim 1$ and $\qi \lesssim 10^{-1}$ in the case of partial aging, and for $\ts \sim 1$ and $\qi \to 0$ in the case of complete aging. More generally, in the complete aging scenario, for any value of $\ts$ the maximum value of~$\ac$ is achieved in the limit $\qi \to 0$. Finally, note that a lack of order, signaled by the vanishing of the critical value~$\ac$, emerges in the limits of $\qi \to 1$ and $\ts \to \infty$ due to the approach to the aging-less case in these limits. Horizontal ($\ac$ vs $\qi$) and vertical ($\ac$ vs $\ts$) slices of the contour plots presented here are shown in Appendix~\ref{app:slices}. 

In Fig.~\ref{fig:taus_qinf>0}, we plot the average age of the agents in each state, namely, 
\begin{equation}
\langle\tau^\pm\rangle=\displaystyle \frac{1}{{x^\pm}}\sum_{\tau}\tau x^\pm_\tau,   
\end{equation}
as a function of noise intensity $a$ for two values of $\ts$. The outcomes from numerical simulations with $N = 10^4$ agents are in good agreement with the theoretical predictions, which are derived in Appendix~\ref{app:Ftau}. 

The mechanism underlying the phase transition exhibited in both the partial and complete aging scenarios can be interpreted as a symmetry breaking related to the average age of the agents in each state. In the disordered phase ($a > \ac$), noise is the dominant update mechanism and no state predominates over the other, resulting in all agents having similar ages regardless of their state. The average ages in the complete aging case are higher than in the partial aging case, as all state changes (whether driven by noise or imitation) are influenced by age, making it less likely for agents to flip their state. In the ordered phase ($a < \ac$), in contrast, there is a preferred opinion and state updates are mainly driven by the imitation mechanism. As a result, agents supporting the majority opinion change state less frequently and have a higher average age compared to those supporting the minority opinion. This leads to an asymmetric distribution of the quantities $\langle \tau^\pm \rangle$, with the one corresponding to the majority opinion predominating over the other. This effect is even more pronounced in the complete aging scenario, as age also influences state updates driven by noise, making it increasingly unlikely for agents supporting the majority opinion to flip their state.

\begin{figure*}[t]
\includegraphics[width=0.4\textwidth]{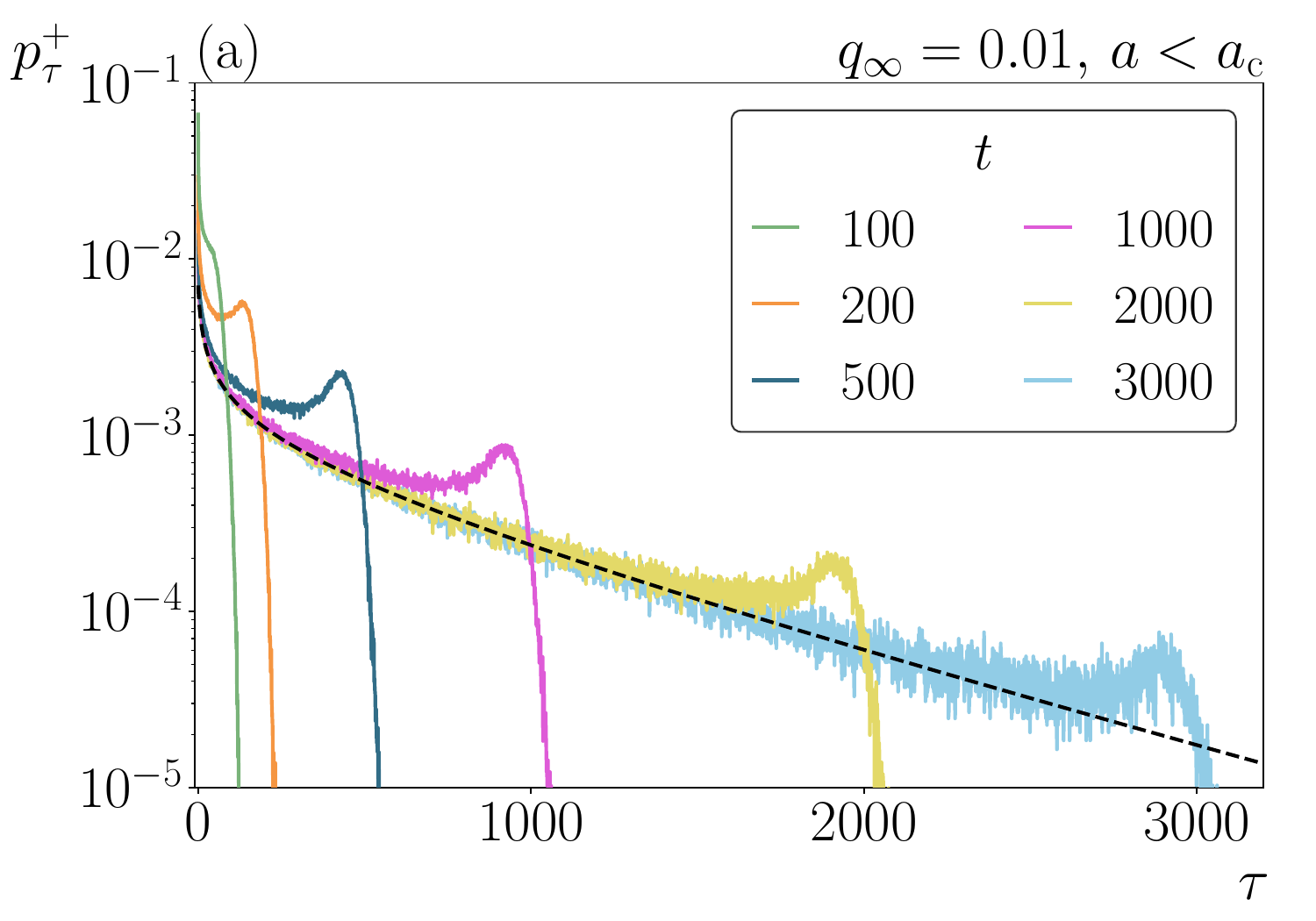}
\includegraphics[width=0.4\textwidth]{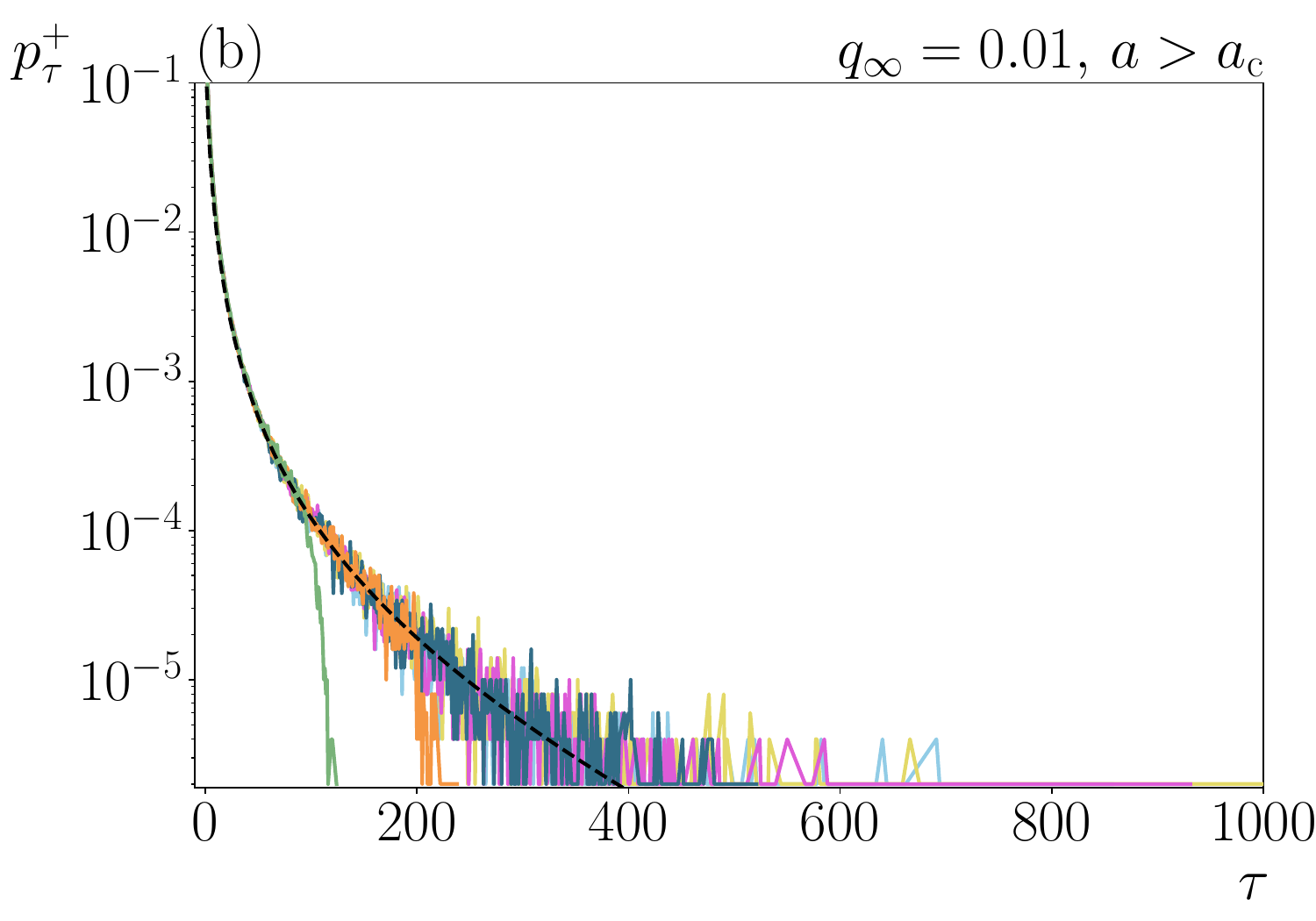}
\includegraphics[width=0.4\textwidth]{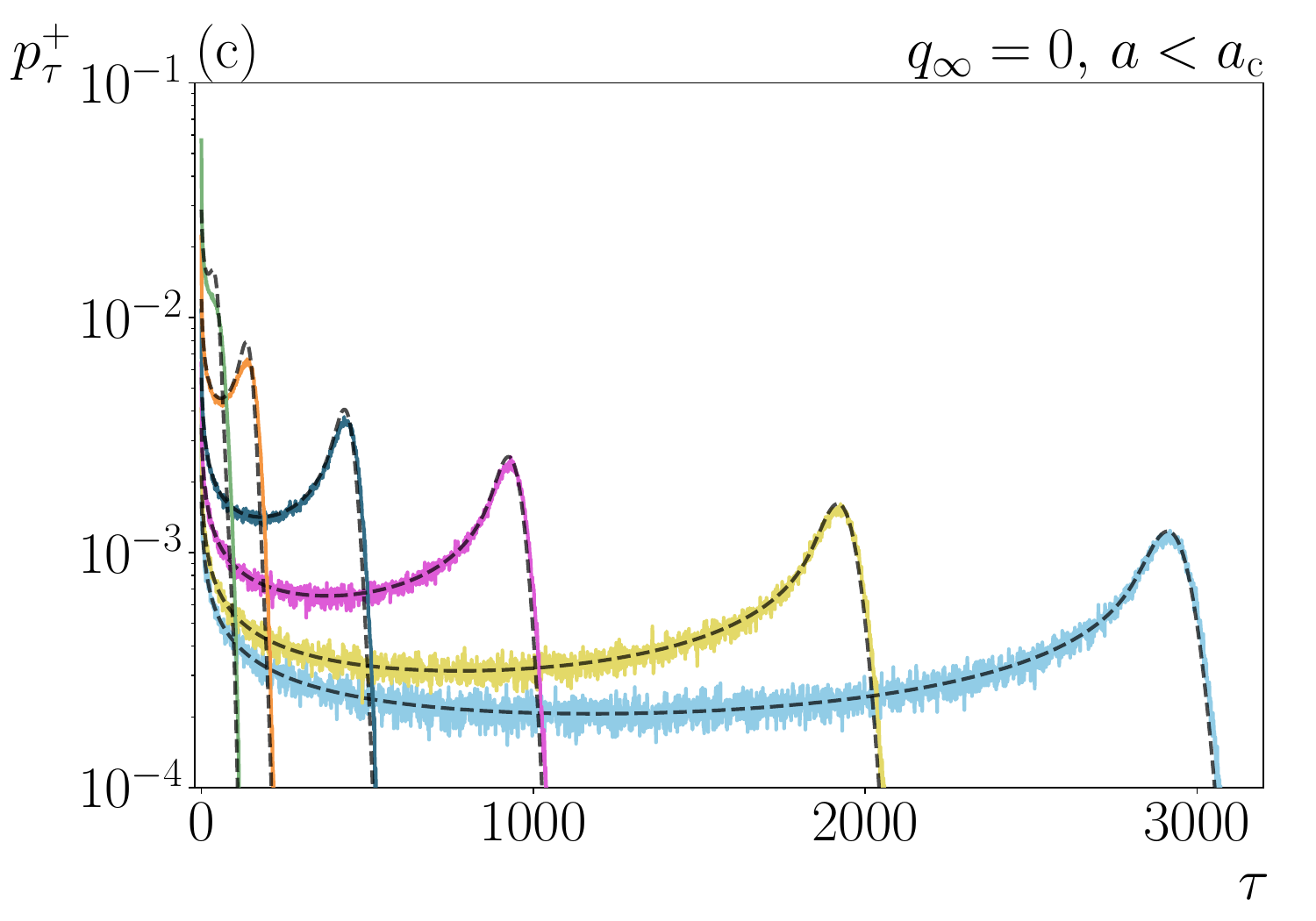}
\includegraphics[width=0.4\textwidth]{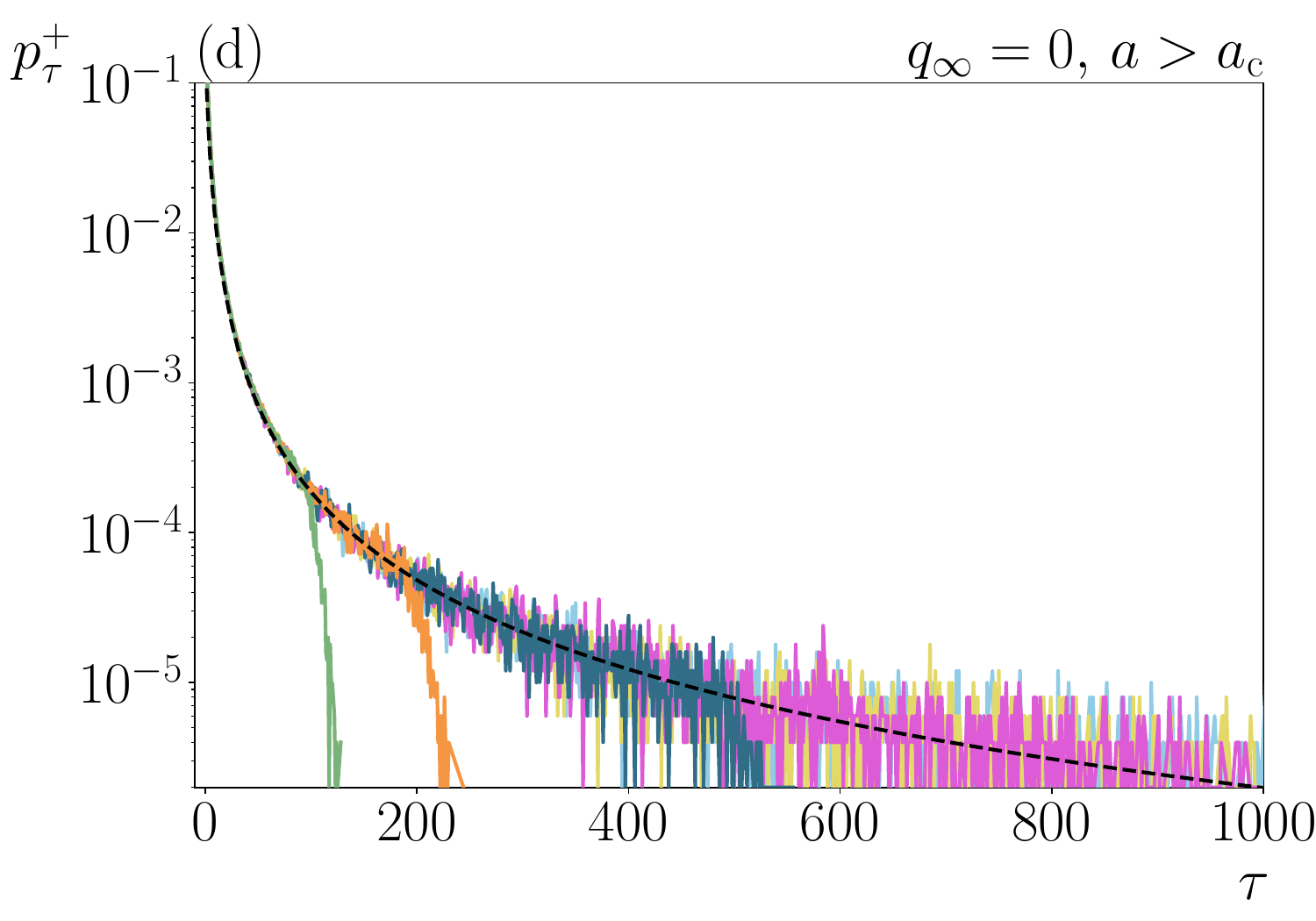}
\caption{Time evolution of the normalized age distribution $p^+_\tau=x_\tau^+/x^+$ in the case of complete aging for $\tau^*=4$ showing behavior below ($a = 0.2$) and above ($a = 0.9$) the critical point $\ac=2/\tau^*$, respectively. Panels (a) and (b) correspond to $q_\infty = 0.01$ while panels (c) and (d) correspond to $q_\infty = 0$. Colored lines correspond to the results of numerical simulations with $N=10^3$ agents averaged over $10^3$ trajectories for different times, as indicated in legend. Black dashed lines represent the mean-field prediction: obtained directly from Eq.~\eqref{eq:app:Vxtau} in (a,b,d), or by numerical integration of the rate equations, Eqs.~(\ref{eq:Vratesi}-\ref{eq:Vratesi0}), in (c). (a,b ) $a=0.2$ and (b,d) $a=0.9$.}
\label{fig:xtau_qinf=0}
\end{figure*}

\subsection{Complete aging for $\qi=0$}
\label{sec:qinf0}
Unlike the partial aging scenario, the complete aging case for $\qi=0$ exhibits different behavior that must be studied separately. The determination of the function $\Phi(x)$ relies on the calculation of the required series that appear in the numerator and denominator of Eqs.~\eqref{eq:Phis}. On the one hand, the series in the numerator, given in Eq.~\eqref{eq:NVM_complete_sumqF}, is convergent for all values of $x,a$ and $\tau^*$. On the other hand, the convergence of the series in the denominator of Eqs.~\eqref{eq:Phis} is not guaranteed. Therefore, the explicit expression given in Eq.~\eqref{eq:NVM_complete_sumF} is not always valid. The Raabe criterion~\cite{chowbook} leads us to the following condition for convergence
\begin{equation} \label{eq:app:Raabe_condition}
 \Lambda(x,1,a)<1-\frac{1}{\tau^*},
\end{equation}
such that
\begin{align} \label{eq:NVM_complete_sumF_qinf=0}
 &\sum_{\tau=0}^{\infty}F_{\tau}(x)=\,_2F_1\left(1,\tau^*\Lambda(x,1,a);\tau^*;1\right)\\ \nonumber
 &=
 \begin{cases} 
 \frac{\displaystyle\tau^*-1}{\displaystyle\tau^*\left[1-\Lambda(x,1,a)\right]-1}, & \text{if $\Lambda(x,1,a)<1-\displaystyle\frac{1}{\tau^*}$} \\
 \infty, & \text{otherwise.} 
 \end{cases}
\end{align}

The convergence condition in Eq.~\eqref{eq:app:Raabe_condition} becomes more intuitive when rewritten as 
\begin{equation} \label{eq:conv-cond} 
x > x^* \equiv \frac{1}{2(1-a)}\left(\frac{2}{\tau^*} - a\right), 
\end{equation} 
which implies that convergence is guaranteed for all values of $x$ only if $x^*<0$. This requirement leads to the condition 
\begin{equation} 
\label{eq:app:ac_qinf=0} 
a > a^* \equiv \frac{2}{\tau^*},
\end{equation}
whereas for $a<a^*$, there exist certain values of $x$ for which the sum diverges and $\Phi(x)$ vanishes. Notably, when $\tau^* < 2$, the convergence condition cannot be satisfied for any $a \in [0,1]$.
 
Focusing on the region $a>a^*$, the function $\Phi(x)$ can be expressed as 
\begin{equation} \label{eq:NVM_complete_Phi_qinf=0}
\Phi(x) = 
 \frac{\displaystyle\tau^*\left[1-\Lambda(x,1,a)\right]-1}{\displaystyle(\tau^*-1)\left[1-\Lambda(x,1,a)\right]}, 
\end{equation}
which reduces Eq.~\eqref{eq:dxdtVgeneral} to the linear equation
\begin{equation}\label{eq:dxdt_qinf=0}
 \frac{dx}{dt}=G(x)=\frac{ \tau^*}{(\tau^*-1)} \left(\frac{2}{\tau^*}-a\right)\left(x- \frac 12\right).
\end{equation}
This equation presents a single fixed point at $\xst=1/2$, which is stable if $a>a^*$. For $a<a^*$, the solution would become unstable, but strictly speaking, Eq.~\eqref{eq:dxdt_qinf=0} is not valid in such region. Nevertheless, the value obtained from the convergence condition, $a^*$, represents the critical value $\ac \equiv a^*$ in the case $\qi=0$. 

The failure of the adiabatic approximation for $a<\ac$ can be understood by analyzing the time evolution of the age distribution $x^+_\tau$, as exemplified in Fig.~\ref{fig:xtau_qinf=0}. The rate equation for the global variable $x$, Eq.~\eqref{eq:dxdtVgeneral}, is derived through an adiabatic elimination, under the assumption that the microscopic variables $x^\pm_\tau$ rapidly reach a stationary state, leaving the dynamics primarily governed by the evolution of the global variable. This approximation is always valid for $\qi>0$, while for $\qi=0$, it holds only if $a>\ac$. In these cases, the age distribution approaches a steady state, and excellent agreement between simulations and theory, given by Eq.~\eqref{eq:app:Vxtau}, is observed. However, for $\qi=0$ and $a<\ac$, the age distribution evolves indefinitely over time, making the adiabatic approximation invalid even though the numerical integration of the rate equations, Eqs.~(\ref{eq:Vratesi}, \ref{eq:Vratesi0}), agrees with the simulation results. 

\begin{figure}[h!] 
\includegraphics[width=0.4\textwidth]{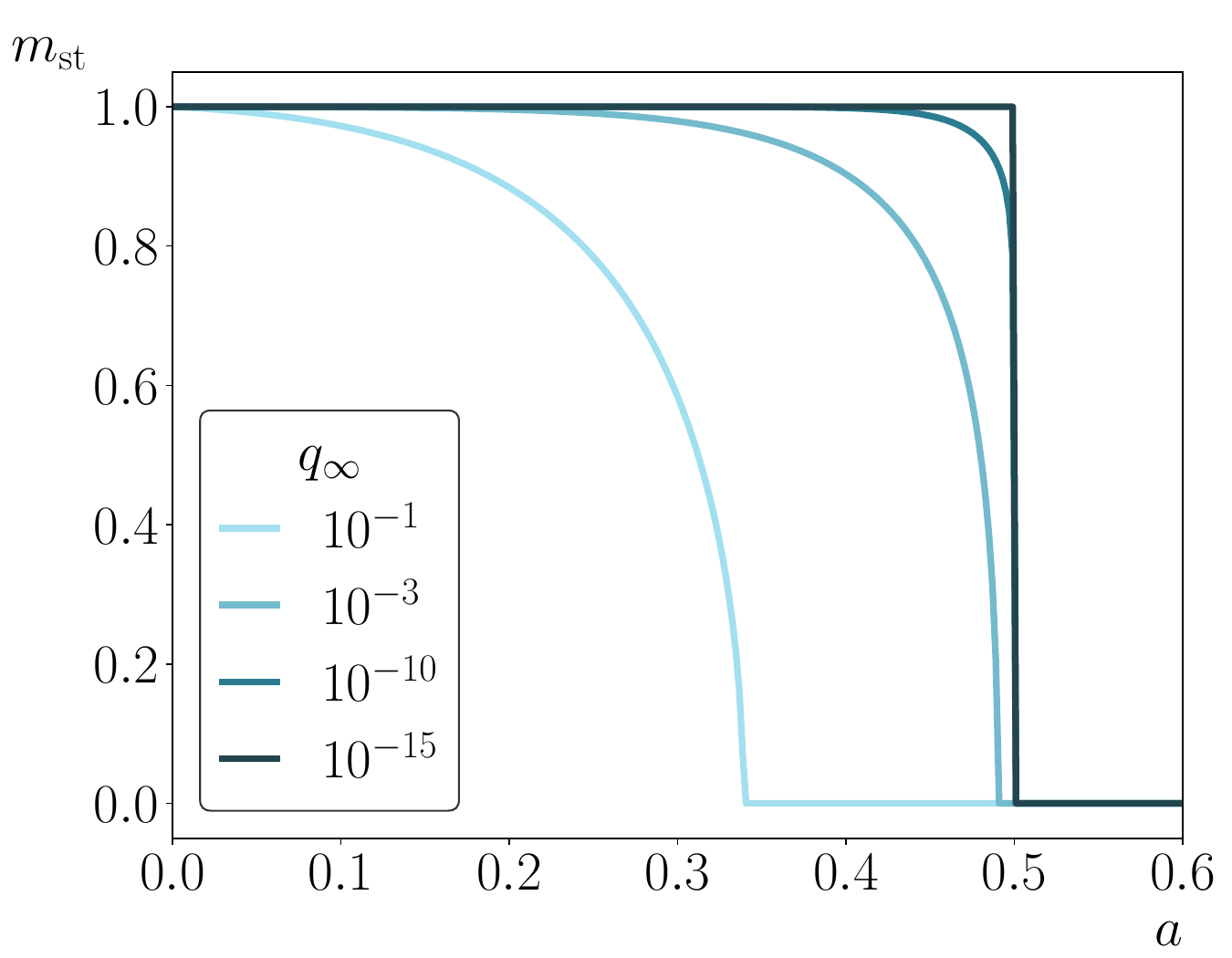}
 \caption{Magnetization, $\mst$, versus the noise intensity, $a$, in the case of complete aging for $\ts = 4$ and several values of $\qi$. The curves represent the numerical solution of the stationary form of Eq.~\eqref{eq:dxdtVgeneral}.} \label{fig:TA_mstvsa_qinfto0}
\end{figure}

This evolution demonstrates a tendency toward an aging population where transitions between states are progressively more difficult. At the beginning of the dynamics, when the agents have zero age, the system rapidly evolves from the initial condition $x = 1/2$ toward one of the two ordered states, say $x=0$. Following this short regime, most agents become older in the state $-1$ and young agents remain more likely to change states. In the voter model, where $a=0$ and only the herding mechanism is activated, the majority drives the system to the absorbing state $x=0$ with an asymptotic power-law behavior $x(t)\sim t^{-\beta}$, $x^+_0(t)\sim t^{-\beta-1}$ with exponent $\beta=\tau^*$~\cite{Peralta2020a}. When considering the idiosyncratic mechanism, $a>0$, agents can change their state independently, which slows the decay toward the ordered state $x=0$. In this case, it is found numerically that both $x(t)$ and $x^+_0(t)$ present the same power-law decay $x(t)\sim x^+_0(t)\sim t^{-\beta}$ with an exponent $\beta>0$ for all the range $a\in(0,\ac)$, see Appendix~\ref{sec:app:xs_temporal}. Random changes of state are increasingly dominant as the system approaches consensus and, since the aging probability is never strictly equal to zero for finite age $\tau$, there is always a chance for the agents to change state even if the system has reached the fully ordered state. Although the theory fails to predict the stable solution $\xst$ in this regime, it can be determined by taking the limit $\qi\to0$ of the stable solutions of Eq.~\eqref{eq:dxdtVgeneral} for $\qi>0$. As $\qi$ decreases, the critical value $\ac(\qi)$ approaches $\ac(\qi = 0) = a^*$ and the second-order transition becomes more abrupt, with the magnetization $\mst$ taking the form of a step function. This phenomenon can be seen in Fig.~\ref{fig:TA_mstvsa_qinfto0} in the case $\ts = 4$, for which $a^* = 0.5$. In the limit $\qi \to 0$, $\mst = 1$ across the entire range $a \in [0, \ac)$. This phenomenon, combined with the decay of $x(t)$ and $x^+_0(t)$, evidences that for all practical purposes a first-order phase transition occurs from $\mst = 1$ to $\mst = 0$ at $a = \ac$. However, this result has yet to be rigorously demonstrated.

This transition can also be understood as a competition between aging effects, characterized by $\tau^*$, and idiosyncratic behavior governed by $a$. For $a<\ac$, the aging probability decays sufficiently fast, driving the system toward an ordered state. On the other hand, for $a>\ac$, the system behaves as if there were no aging present, rapidly reaching the disordered state.

\section{Conclusions}
\label{sec:conclusions}

Previous studies on aging in the literature have exclusively considered memory effects that influence the herding mechanism. In this paper, we extend the analysis of the noisy voter model with aging by also incorporating the influence of memory effects on idiosyncratic changes. We have derived a mean-field description for generic forms of aging kernels that independently influence each mechanism of interaction, allowing for potential future extensions. However, our study specifically focuses on the case where the same algebraic aging governs both mechanisms. 

The complete aging scenario with $\qi>0$ is qualitatively similar to the partial case, with a second-order phase transition occurring for a critical value of the noise intensity, $\ac$. However, compared to the partial aging scenario, the consensus region expands due to a shift of the critical value $\ac$ to higher values. Additionally, the asymmetry between the average ages of agents in each state, $\langle\tau^\pm\rangle$ becomes more pronounced.

In contrast, the complete aging case exhibits a unique regime for $\qi=0$. The adiabatic approximation becomes invalid for $a<\ac$, since the microscopic variables~$x^\pm_\tau$ do not reach a steady state, evolving indefinitely over time. However, numerical integration of the mean-field equations still allows for accurate prediction of the results from numerical simulations. The system evolves toward one of the ordered states exhibiting a power-law decay over time. Although not rigorously demonstrated, for all purposes, a first-order transition from a completely ordered to a disordered state for $a=\ac$ is observed.

Exploring alternative forms of the aging kernel could reveal new features. In our previous work on the noisy voter model with partial aging~\cite{ourPhysics2024}, we did not observe qualitatively different outcomes for algebraic and exponential decays of the aging probability $q(\tau)$. However, in the context of complete aging, the power-law decay $q(\tau)\sim \tau^{-1}$ may represent a marginal case under certain conditions, as suggested by previous work in the voter model~\cite{Peralta2020a}. 
Preliminary analysis indicates that, for $q_\infty\neq0$, the system orders regardless of the decay rate of the aging probability, while for $q_\infty=0$, the behavior depends on the law of decay. 
We conjecture that, for decays faster than $1/\tau$, the system orders in the same manner as in the case $q_\infty\neq 0$, while for slower decays, the system becomes trapped in a frozen state. This conjecture was verified for specific cases such $q_\tau\sim 1/\tau^{1/2}$, where order occurs, and for the exponential decay case, $q_\tau\sim e^{-\tau/\tau^*}$, where the system quickly reaches a frozen state, determined by both the initial condition $x(0)$ and the noise intensity $a$. 
This issue requires further extensive investigation, which we leave for future work. 

Another promising direction for future research would be to move beyond all-to-all interactions and explore the effects of complete aging on opinion dynamics in random networks. 

\acknowledgments{
Partial financial support has been received from the Agencia Estatal de Investigaci\'on (AEI, MCI, Spain) MCIN/AEI/10.13039/501100011033 and Fondo Europeo de Desarrollo Regional (FEDER, UE) under Project APASOS (PID2021-122256NB-C21) and the María de Maeztu Program for units of Excellence in R\&D, grant CEX2021-001164-M. 
C.A. acknowledges partial support received from
 Conselho Nacional de Desenvolvimento Científico e Tecnológico (CNPq)-Brazil (311435/2020-3), Fundação de Amparo à Pesquisa do
Estado de Rio de Janeiro (FAPERJ)-Brazil (CNE E-26/204.130/2024), and Coordenação de Aperfeiçoamento de Pessoal de Nível Superior 
(CAPES)-Brazil (code 001). S.O-B. acknowledges support from the Spanish Ministry of Education and Professional Training under the grant FPU21/04997.
}

\bibliographystyle{unsrt}
\bibliography{total_aging}

\appendix
\setcounter{section}{0}
\setcounter{equation}{0}
\widetext
\section{Rate equations}
\label{app:rates}
Let us denote by $x^s_\tau$ the fraction of agents in state $s=\pm1$ and age $\tau=0,1,\dots$, 
their corresponding rate equations are given by
\begin{equation} \label{eq:app:dxdt_tau}
 \frac{dx_{\tau}^{s}}{dt}=\Omega_{ss}(\tau-1)-\sum_{s'}{\Omega_{ss'}(\tau)},\quad \tau\ge 1,
\end{equation}
where $\Omega_{ss'}(\tau)\equiv\Omega(s\to s',\tau)$ is the transition rate from state $s$ to state $s'$ for an agent with age $\tau$. The first term represents the updates for agents with age $\tau-1$ whose states did not change, leading to an increase in their internal time by one unit, i.e., $\tau-1 \to \tau$. The second term accounts for updates of agents with age $\tau$, either those whose state changed, resetting their internal time to zero (i.e., $\tau \to 0$) or those whose state remained unchanged, resulting in an increase of their internal time by one unit, $\tau \to \tau+1$.

The case of agents with age $\tau=0$, requires an special treatment, giving the following rate equation
\begin{equation} \label{eq:app:dxdt_0}
 \frac{dx_{0}^{s}}{dt}=\sum_{\tau=0}^{\infty}\sum_{s'\neq s}{\Omega_{s's}(\tau)}-\sum_{s'}{\Omega_{ss'}(0)},
\end{equation}
where the first term accounts for agents, regardless of their age, that have changed their state and consequently its internal time is reset to zero, $\tau \to 0$, and the second term is analogous to the one of Eq.~\eqref{eq:app:dxdt_tau}.

Summing Eqs.~(\ref{eq:app:dxdt_tau}, \ref{eq:app:dxdt_0}) over all values of $\tau$, one obtains the rate equations for the fraction of agents in state $s$ as $x^s=\sum_{\tau=0}^{\infty}x^s_\tau$, which reads
\begin{equation}
 \frac{dx^{s}}{dt}=\sum_{\tau=0}^{\infty}\sum_{s'}\biggl({\Omega_{s's}(\tau)}-{\Omega_{ss'}(\tau)}\biggr).
\end{equation}

In order to derive the particular rate equations of our model, let us explain in detail the rate $\Omega_{--}(\tau)$, for which an agent with state $-1$ and age $\tau$ has been selected and does not change its state. First, it has to be selected with probability $x^-_\tau$. If the idiosyncratic rule is selected, with probability $a$, the agent keeps its state if either the mechanism is not activated, with probability $1-\tilde{q}_\tau$, or if the rule is activated, with probability $\tilde{q}_\tau$, and the state $-1$ is selected at random, with probability $1/2$. If, on the other hand, the social rule is chosen, with probability $1-a$, the agent remains in the same state if either the copying mechanism is not activated, with probability $1-q_\tau$, or if the copying mechanism is activated, with probability $q_\tau$, and a neighbor with state $-1$ is chosen, with probability $x^-$. These possible scenarios lead to the rate
\begin{equation}
\Omega_{--}(\tau)=x^-_\tau\left[a \left(\frac{\tilde{q}_\tau}{2}+(1-\tilde{q}_\tau)\right)+(1-a)\left(q_\tau x^-+(1-q_\tau)\right)\right].
\end{equation}
The other rates $\Omega_{ss'}$ can be calculated with similar reasoning. After simplification, the final result is
 \begin{equation}
 \label{eq:Vomegas}
 \begin{split}
\Omega_{--}(\tau)&= x_{\tau}^{-}\biggl[\frac{a}{2}(2-\tilde{q}_\tau) +(1-a)(1-x^+ q_\tau) \biggr], 
\\
\Omega_{-+}(\tau)&= x_{\tau}^{-}\biggl[\frac{a}{2} \tilde{q}_\tau +(1-a)x^+ q_\tau \biggr],
\\
\Omega_{+-}(\tau)&= x_{\tau}^{+}\biggl[\frac{a}{2}\tilde{q}_\tau +(1-a)x^-q_\tau \biggr],
\\
\Omega_{++}(\tau) &= x_{\tau}^{+}\biggl[\frac{a}{2}(2-\tilde{q}_\tau) +(1-a)(1-x^-q_\tau)\biggr] ,
\end{split}
\end{equation}
where the relation $x^-=1-x^+$ has been used for convenience.

From these rates, one can write Eqs.~(\ref{eq:app:dxdt_tau}, \ref{eq:app:dxdt_0}) as
\begin{equation} \label{eq:Vratesi}
 \begin{split}
 \frac{dx_\tau^-}{dt} &=\Omega_{--}(\tau-1)-x_\tau^-,\\
 \frac{dx_\tau^+}{dt} &=\Omega_{++}(\tau-1)-x_\tau^+,
\textbf{} \end{split}
\end{equation}
for $\tau\geq1$, and
\begin{equation} \label{eq:Vratesi0}
 \begin{split}
 \frac{d x^{-}_0}{dt} &
 = \frac{a}{2} \tilde{y}^+ +(1-a)x^-y^+ -x_0^-,
\\
\frac{d x^{+}_0}{dt} &
 = \frac{a}{2} \tilde{y}^- +(1-a)x^+y^- -x_0^+,
 \end{split}
\end{equation}
for $\tau=0$, where we have defined 
\begin{equation} \label{eq:app:ys}
 y^s \equiv \sum_{\tau=0}^{\infty} q_\tau x^s_\tau,\;\;\;\;
 \tilde{y}^s \equiv \sum_{\tau=0}^{\infty} \tilde{q}_\tau x^s_\tau.
\end{equation}
With the aim of obtaining a closed equation for the time evolution of the global variables $x^s$, we use an adiabatic approximation whereby we assume that the microscopic variables $x_\tau^s$ rapidly arrive to the stationary state and the time derivatives of Eqs.~(\ref{eq:Vratesi},\ref{eq:Vratesi0}) can be set to zero. We remark that for complete aging and $\qi=0$, this assumption does not always hold, as discussed in Sec.~\ref{sec:qinf0}. When valid, it leads to the following recursive relation
\begin{equation} \label{eq:xtau}
\begin{split}
x_\tau^-&=x_0^- F_\tau(x^+),\\
x_\tau^+&=x_0^+ F_\tau(x^-),
\end{split}
\end{equation}
with
\begin{equation} \label{app:eq:Ftau}
F_0(x)=1, \hspace{1cm}
F_\tau(x)\equiv \prod_{k=0}^{\tau-1}\Lambda(q_k\, x,\tilde{q}_k,a),\quad \tau\ge 1,
\end{equation}
and
\begin{equation} \label{app:eq:Lambda_def}
\Lambda(z,q,a)= a\left(1-\frac{q}{2} \right) +(1-a)(1-z).
\end{equation}
Summing Eq.~(\ref{eq:xtau}) over all $\tau\ge 0$, we obtain
\begin{equation} \label{eq:xxx}
\begin{split}
x^- &= x_0^-\sum_{\tau=0}^{\infty}F_\tau(x^+),\\ 
x^+ &= x_0^+\sum_{\tau=0}^{\infty}F_\tau(x^-),
\end{split}
\end{equation}
which substituted in Eqs.~\eqref{eq:xtau} leads to
\begin{equation} \label{eq:app:Vxtau}
\begin{split}
x_\tau^-&=x^- \frac{F_\tau(x^+)}{\sum_\tau F_\tau(x^+)},\\
x_\tau^+&=x^+ \frac{F_\tau(x^-)}{\sum_\tau F_\tau(x^-)},
\end{split}
\end{equation}
which are now expressed in terms of the global variables $x^\pm$.

On the other hand, adding Eqs.~(\ref{eq:Vratesi}) and (\ref{eq:Vratesi0}) over all values of $\tau$, we obtain the corresponding equations for the fraction $x^s$ of each state $s$. 
\begin{equation} \label{eq:app:Vrates+-}
 \begin{split}
\frac{d x^{-}}{dt} &=
\frac{a}{2}( \tilde{y}^+ -\tilde{y}^-) 
+ (1-a)( x^-y^+ -x^+ y^-) , \\
\frac{d x^{+}}{dt} &=
\frac{a}{2}( \tilde{y}^- -\tilde{y}^+) 
+ (1-a)(x^+ y^- - x^- y^+).
\end{split}
\end{equation}

Moreover, note that one of the two equations can be eliminated since $x^+ +x^- =1$. 
Then, to obtain a closed evolution equation for the global variable $x^+$, we need to express the variables $y^s$ and $\tilde{y}^s$ appearing in Eqs.~\eqref{eq:app:Vrates+-} in terms of $x^+$. 
This can be done by using Eq.~(\ref{eq:app:Vxtau}) into Eqs.~\eqref{eq:app:ys}, yielding
\begin{equation} \label{eq:Vyyy}
\begin{split}
y^- =(1-x^+) \Phi(x^+), \;\;\;\;\; y^+ &=x^+ \Phi(1-x^+), \\
 \tilde{y}^- = (1-x^+) \tilde\Phi(x^+), \;\;\;\;\;
\tilde{y}^+ &=x^+ \tilde\Phi(1-x^+), 
\end{split}
\end{equation}
where we have introduced the functions
\begin{equation} \label{app:eq:Phis}
\Phi(x) \equiv 
\frac{\sum_{\tau=0}^\infty q_\tau F_\tau(x)}{\sum_{\tau=0}^\infty F_\tau(x)} 
\;\;\;\mbox{and}\;\;\; 
\tilde\Phi(x) \equiv 
\frac{\sum_{\tau=0}^\infty \tilde{q}_\tau F_\tau(x)}{\sum_{\tau=0}^\infty F_\tau(x)} 
.
\end{equation}
Let us note here, for consistency, that in the aging-less case, $\tilde{q}_\tau=q_\tau=1$, it is $\tilde\Phi(x)=\Phi(x)= 1$ and, hence, $\tilde{y}^s=y^s=x^s$, for $s=\pm1$. 

Replacement of Eqs.~\eqref{eq:Vyyy} in Eqs.~\eqref{eq:app:Vrates+-} leads to the following rate equation for $x\equiv x^+$
\begin{align}
\frac{dx}{dt}=G(x)\equiv \frac{a}{2} \left[(1-x)\tilde\Phi(x) - x\tilde\Phi(1-x)\right] + (1-a) x (1-x) \left[\Phi(x)- \Phi(1-x)\right].
\end{align}

\section{Calculation of sums involving $F_\tau(x)$}
\label{app:Ftau}
Let us assume that the herding and the idiosyncratic mechanisms are modeled by the following rational function of the age 
\begin{equation}
 q_{\tau}=\frac{q_{\infty}\tau+ \tau^*}{\tau+\tau^*}, \, \tilde{q}_{\tau}=\frac{\tilde{q}_{\infty}\tau+ \tau^*}{\tau+\tau^*}, 
\end{equation} 
respectively, where $\tau^*>0$ and $ q_{\infty},\,\tilde{q}_{\infty} \in [0,1]$. The function $F_\tau(x)$ defined in Eq.~\eqref{eq:Ftau} is given by
\begin{equation}
F_\tau(x)\equiv \prod_{k=0}^{\tau-1}\Lambda(q_k\, x,\tilde{q}_k,a)=\Lambda(\qi x,\tilde{q}_\infty,a)^{\tau }\frac{ \left(\tau^*\xi(x,a,\qi,\tilde{q}_\infty)\right)_{\tau }}{(\tau^*)_{\tau }},\quad \tau\ge 1,
\end{equation}
where $(z)_\tau\equiv \Gamma(z+\tau)/\Gamma(z)$ is the Pochhammer symbol, and the function $\xi(x,a, \qi, \tilde{q}_\infty)$ reads

\begin{equation}
\xi(x,a,\qi,\tilde{q}_\infty)\equiv\frac{\Lambda( x, 1,a)}{\Lambda(\qi x, \tilde{q}_\infty,a)},
\end{equation}
with
\begin{equation}
 \Lambda(x, q,a)= a\left(1-q/2 \right) +(1-a)(1-x), 
\end{equation} 
as defined in Eq.~\eqref{eq:Lambda_def}.

The sums required to calculate $\Phi(x)$, $\tilde{\Phi}(x)$ 
as defined in Eqs.~(\ref{app:eq:Phis}), take the following explicit forms in terms of the hypergeometric function $_2F_1(a,b;c;z)$,
\begin{align} 
 \sum_{\tau=0}^{\infty}F_{\tau}(x)
 &=\,_2F_1\left(1,\tau^*\xi(x,a,\qi,\tilde{q}_\infty);\tau^*;\Lambda(\qi x, \tilde{q}_\infty,a) \right), \label{eq:NVM_complete_sumF}\\
 \sum_{\tau=0}^{\infty}q_{\tau}F_{\tau}(x)&= \,_2F_1\left(1,\tau^*\xi(x,a,\qi,\tilde{q}_\infty);1+\tau^*;\Lambda(\qi x, \tilde{q}_\infty,a) \right) \nonumber\\
 &+\frac{q_{\infty} }{1+\tau^*}\,\Lambda( x, 1,a)\, _2F_1\left(2,1+\tau^*\xi(x,a,\qi,\tilde{q}_\infty);2+\tau^*;\Lambda(\qi x, \tilde{q}_\infty,a)\right), \label{eq:NVM_complete_sumqF}\\
 \sum_{\tau=0}^{\infty}\tilde{q}_{\tau}F_{\tau}(x)&= \,_2F_1\left(1,\tau^*\xi(x,a,\qi,\tilde{q}_\infty);1+\tau^*;\Lambda(\qi x, \tilde{q}_\infty,a) \right) \nonumber\\
 &+\frac{\tilde{q}_{\infty} }{1+\tau^*}\,\Lambda( x, 1,a)\, _2F_1\left(2,1+\tau^*\xi(x,a,\qi,\tilde{q}_\infty);2+\tau^*;\Lambda(\qi x, \tilde{q}_\infty,a)\right). \label{eq:NVM_complete_sumqtF}
\end{align}
In the particular case of complete aging, where $q_\tau=\tilde{q}_\tau$, Eqs.~(\ref{eq:NVM_complete_sumqF},\ref{eq:NVM_complete_sumqtF}) become \begin{equation} \label{eq:NVM_complete_sumqF_common}
 \sum_{\tau=0}^{\infty}q_{\tau}F_{\tau}(x)=\frac{1}{1-\Lambda(x,1,a)},
\end{equation}
irrespective of the parameters $\tau^*,\qi$.

Back to the general case, and in order to compute the average age of agents in each state $\langle\tau^{\pm}\rangle=\displaystyle \frac{1}{{x^\pm}}\sum_{\tau}\tau x^\pm_\tau$, we use Eq.~\eqref{eq:app:Vxtau} and calculate the following sum
\begin{equation}
 \sum_{\tau=0}^{\infty}\tau F_{\tau}(x)=\Lambda( x,1, a)\,_2F_1\left(2,1+\tau^*\xi(x,a,\qi,\tilde{q}_\infty);1+\tau^*;\Lambda(\qi\,x, \tilde{q}_\infty, a) \right),
\end{equation}
such that
\begin{equation} \label{eq:app:mean_taus}
 \langle\tau^{\pm}\rangle =\mathcal{T}\left(x^{\mp},\qi,\tilde{q}_{\infty},a\right), 
\end{equation}
where $x^+\equiv x$ and $x^-\equiv 1-x$ are the stable solutions of Eq.~\eqref{eq:dxdtVgeneral} and the function $\mathcal{T}$ is given by
\begin{equation} \label{eq:app:general_T}
\mathcal{T}\left(x,\qi, \tilde{q}_{\infty},a\right)=\Lambda(x,1,a)\,\frac{_2F_1\left(2,1+\tau^*\xi(x,a,\qi,\tilde{q}_\infty);1+\tau^*;\Lambda(\qi\,x,\tilde{q}_\infty,a)\right)}{_2F_1\left(1,\tau^*\xi(x,a,\qi,\tilde{q}_\infty);\tau^*;\Lambda(\qi\,x,\tilde{q}_\infty ,a) \right)}.
\end{equation}

\newpage
\section{Slices of the contour plots of the critical point $\ac$ in the $(\qi, \ts)$ plane}\label{app:slices}

\begin{figure}[h!] \includegraphics[width=0.4\textwidth]{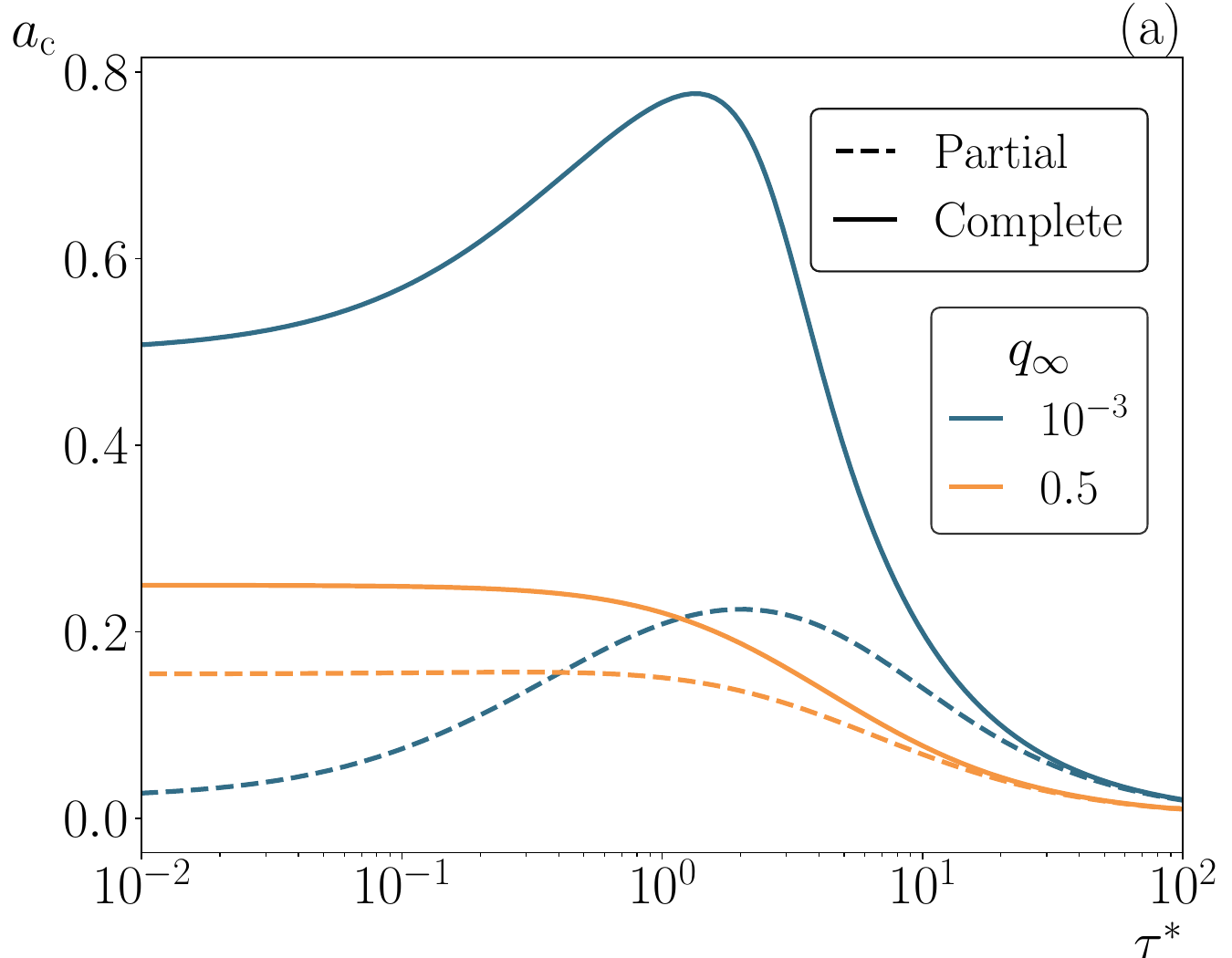}
\includegraphics[width=0.4\textwidth]{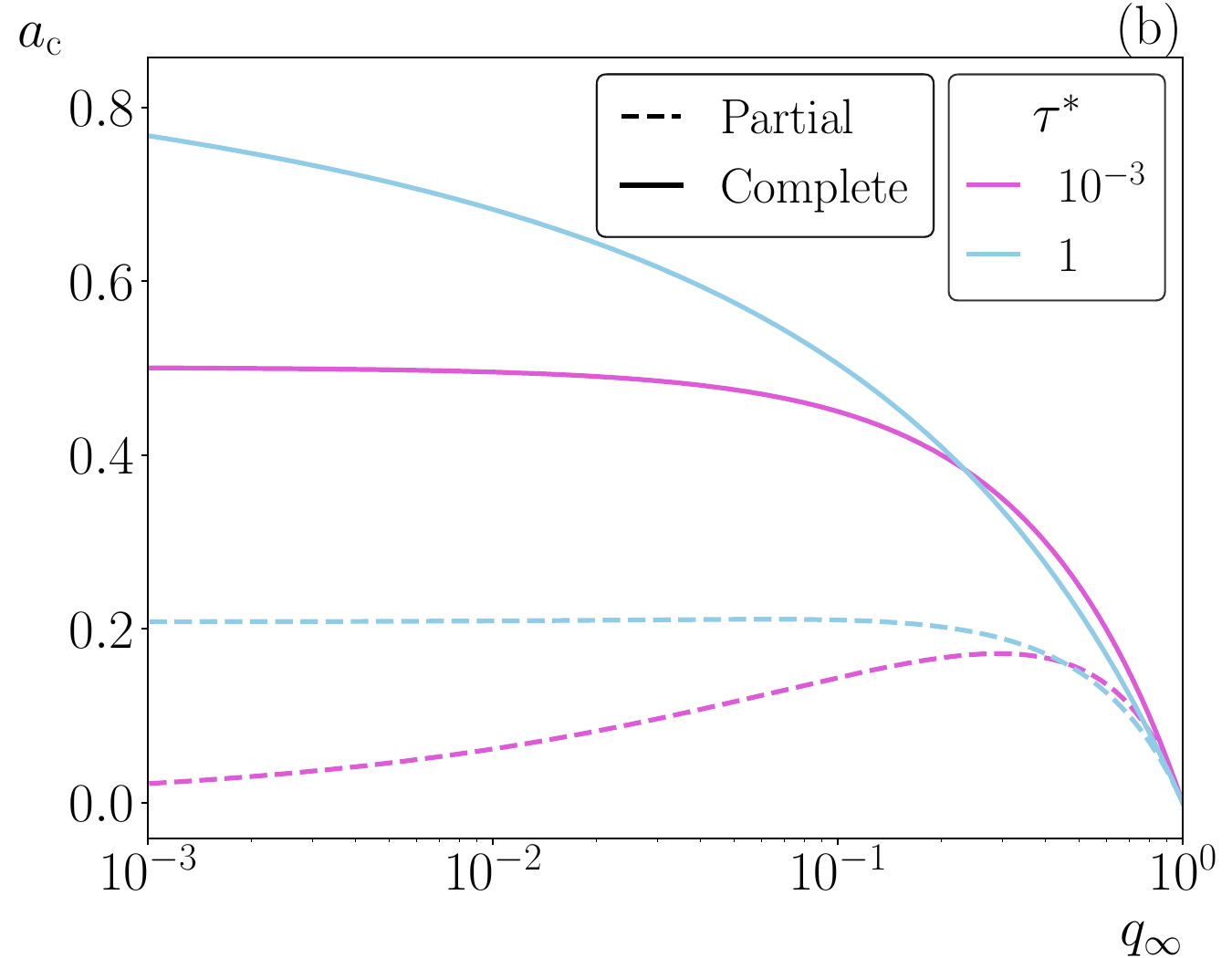}
 \caption{Vertical (a) and horizontal (b) cuts of the heat plots of Fig.~\ref{fig:acs_heat}. Critical values $\ac$ versus $\tau^*$ (a) and $\qi$ (b), for partial and complete aging.}
 \label{fig:acs_slice}
\end{figure}

In Fig.~\ref{fig:acs_slice}, we show horizontal and vertical slices of the contour plots of the critical value $\ac$ in the parameter space $(\qi, \ts)$ presented in Fig.~\ref{fig:acs_heat}. We note that the main effect of complete aging is to increase the value of the critical point $\ac$, leading to larger consensus regions compared to the partial aging case.

\section{Time evolution of the global variable for $\qi=0$} \label{sec:app:xs_temporal}
\begin{figure}[h!] 
\includegraphics[width=0.45\textwidth]{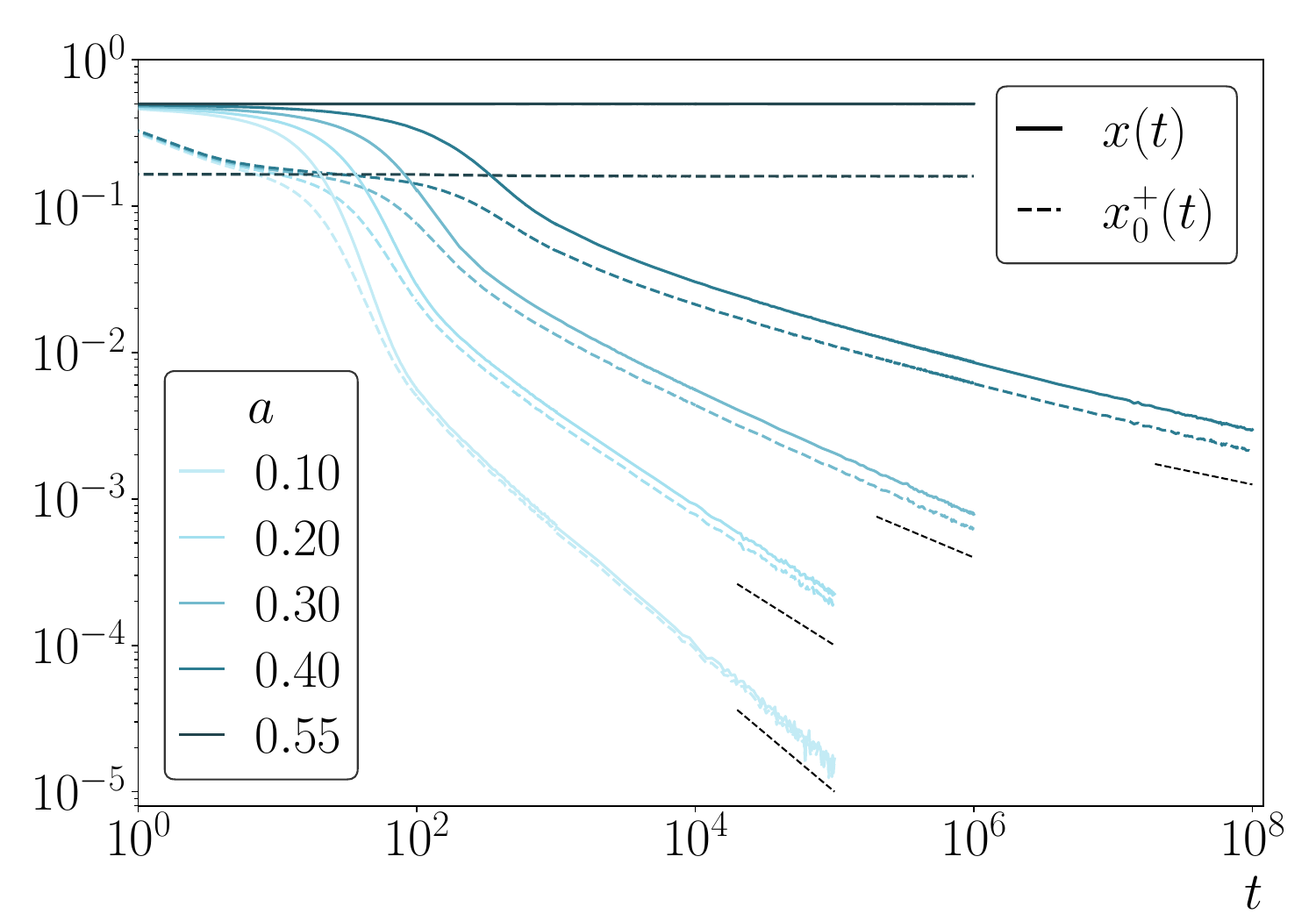}
\includegraphics[width=0.45\textwidth]{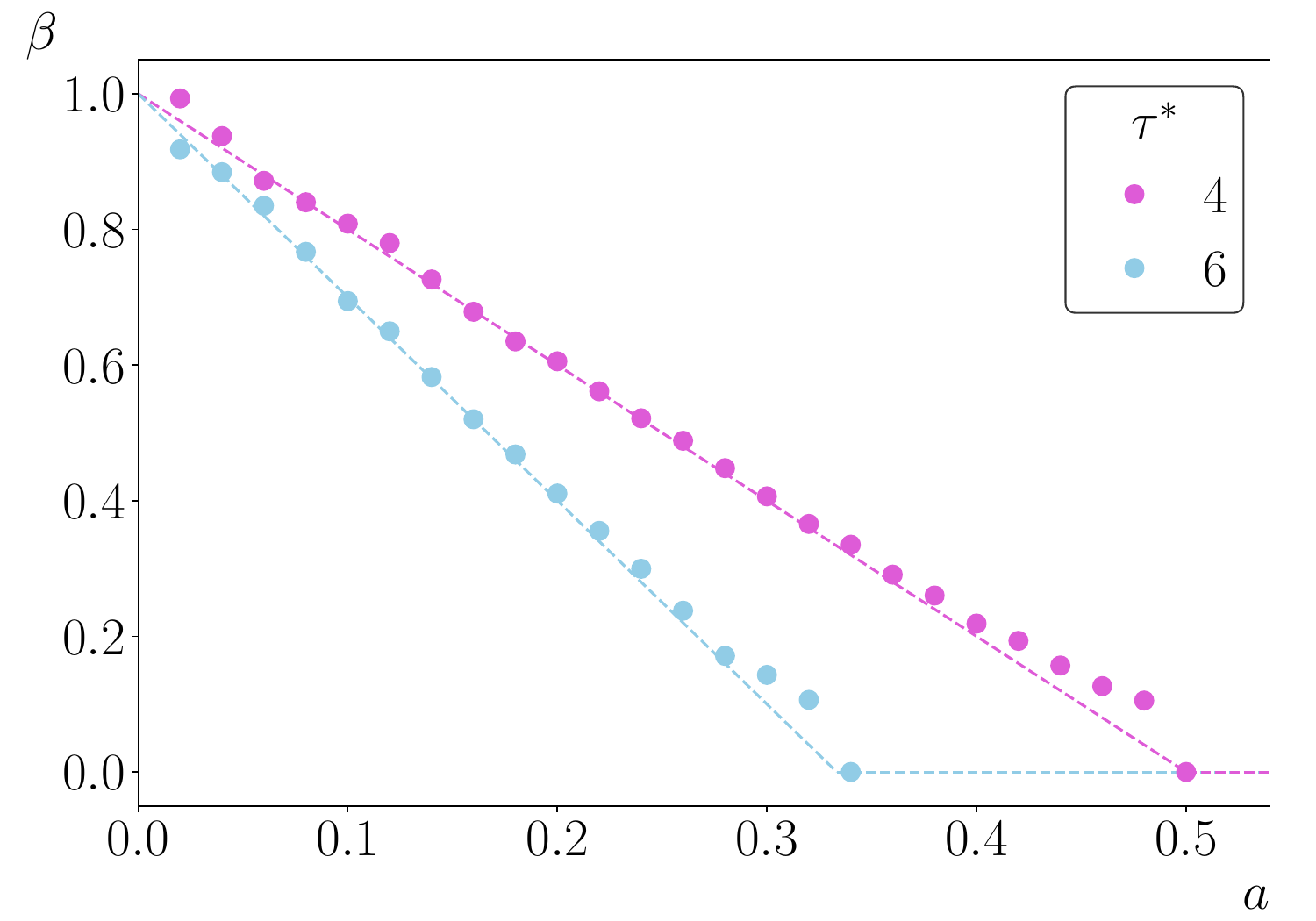}
 \caption{(a) Time evolution of the densities $x(t)$ and $x_0^+(t)$ for different values of $a$ indicated in the legend for $\qi=0, \tau^*=4$. The segments have slope $\beta$, obtained by means of a linear fit to the asymptotic region of each curve. (b) Exponent $\beta$ versus the noise intensity $a$, for values $\tau^*=4,6$. Dashed lines correspond to a fitting of the form $\beta=1-a\tau^* /2$.}
 \label{fig:app:xs_temporal}
\end{figure}

Below the critical point, the microscopic variables $x_\tau^\pm$ do not reach a stationary state, evolving indefinitely over time and driving the global state of the system $x$ toward one of the ordered states, say $x=0$. After a transient period, a power-law decay of the form $t^{-\beta}$ is observed for both $x(t)$ and $x_0^+(t)$ (see Fig.~\ref{fig:app:xs_temporal}). The closer the system is to the critical point, the longer the transient time. The exponent $\beta$ is positive for all the range $a\in(0,\ac)$ and decreases as the system approaches the critical point $\ac$.

\end{document}